\documentclass[twocolumn]{openjournal-bjm-accepted}
\usepackage{natbib}
\usepackage{amssymb}
\usepackage{aas_macros}
\usepackage{url}
\usepackage{amsmath}
\usepackage{fixltx2e}

\usepackage{amsmath}

\newcommand{\lik}{\ensuremath{{\cal L}}}

\newcommand{\Chandra}{\emph{Chandra}}

\newcommand{\XMM}{\emph{XMM-Newton}}

\newcommand{\chisq}{\ensuremath{\chi^2}}

\newcommand{\gta}{\,\rlap{\raise 0.4ex\hbox{$>$}}{\lower 0.6ex\hbox{$\sim$}}\,}
\newcommand{\lta}{\,\rlap{\raise 0.4ex\hbox{$<$}}{\lower 0.6ex\hbox{$\sim$}}\,}
\newcommand{\diff}{\text{d}}

\newcommand{\cm}{\mbox{\ensuremath{\text{~cm}}}}

\newcommand{\s}{\mbox{\ensuremath{\text{~s}}}}
\newcommand{\ks}{\mbox{\ensuremath{\text{~ks}}}}

\newcommand{\keV}{\mbox{\ensuremath{\text{~keV}}}}

\newcommand{\erg}{\mbox{\ensuremath{\text{~erg}}}}

\newcommand{\degree}{\ensuremath{\mathrm{^\circ}}}

\newcommand{\pcmsq}{\ensuremath{\cm^{-2}}}

\newcommand{\ps}{\ensuremath{\s^{-1}}}

\newcommand{\flux}{\ensuremath{\erg \ps \pcmsq}}

\newcommand{\apropto}{\mathrel{\vcenter{
  \offinterlineskip\halign{\hfil$##$\cr
    \propto\cr\noalign{\kern2pt}\sim\cr\noalign{\kern-2pt}}}}}

\newcommand{\lognlogs}{\ensuremath{\log(N)-\log(S)}}

\newcommand{\dpois}{\ensuremath{\text{dpois}}}
\newcommand{\dlnorm}{\ensuremath{\text{dlnorm}}}

\usepackage[utf8]{inputenc}
\usepackage[T1]{fontenc}
\usepackage{graphicx}
\usepackage{grffile}
\usepackage{longtable}
\usepackage{wrapfig}
\usepackage{rotating}
\usepackage[normalem]{ulem}
\usepackage{amsmath}
\usepackage{textcomp}
\usepackage{amssymb}
\usepackage{capt-of}
\usepackage{hyperref}
\usepackage{enumitem}
\setitemize{noitemsep,topsep=5pt,parsep=5pt,partopsep=0pt,leftmargin=10pt}
\usepackage{setspace}
\usepackage{listings}
\usepackage[dvipsnames]{xcolor}

\usepackage{rcs}
\RCS $Date: 2019/02/18 20:58:22 $
\RCS $Revision: 1.7 $
\hypersetup{
 pdfauthor={Ben Maughan},
 pdftitle={},
 pdfkeywords={},
 pdfsubject={},
 pdfcreator={Emacs 26.1 (Org mode 9.1.14)},
 pdflang={English}}
\begin{document}

\title{The X-ray Variability of AGN and its Implications for Observations of Galaxy Clusters.}
\shorttitle{X-ray Variability of AGN}
\author{Ben J. Maughan}
\affil{H. H. Wills Physics Laboratory, University of Bristol, Tyndall Ave, Bristol BS8 1TL, UK}
\email{ben.maughan@bristol.ac.uk}
\and
\author{Thomas H. Reiprich}
\affil{Argelander-Institut für Astronomie, Universität Bonn, Auf dem Hügel 71, 53121 Bonn, Germany}

\begin{abstract}
The detection of new clusters of galaxies or the study of known
clusters of galaxies in X-rays can be complicated by the presence of
X-ray point sources, the majority of which will be active galactic
nuclei (AGN). This can be addressed by combining observations from a
high angular resolution observatory (such as \(\Chandra\)) with deeper
data from an observatory with a larger collecting area, but that may
not be able to resolve the AGN (like \(\XMM\)). However, this approach
is undermined if the AGN varies in flux between the epochs of the
observations. To address this we measure the characteristic X-ray
variability of serendipitously detected AGN in 70 pairs of \(\Chandra\)
observations, separated by intervals of between one month and thirteen
years. After quality cuts, the full sample consists of 1511 sources,
although the main analysis uses a subset of 416 sources selected on
the geometric mean of their flux in the pairs of observations, which
eliminates selection biases. We find a fractional variability that
increases with increasing interval between observations, from about
\(0.25\) for observations separated by tens of days up to about \(0.45\)
for observations separated by \(\sim 10\) years. As a rule of thumb,
given the precise X-ray flux of a typical AGN at one epoch, its flux
at a second epoch some years earlier or later can be predicted with a
precision of about \(60\%\) due to its variability (ignoring any
statistical noise). This is larger than the characteristic variability
of the population by a factor of \(\sqrt{2}\) due to the uncertainty on
the mean flux of the AGN due to a single prior measurement. The
precision can thus be improved with multiple prior flux measurements
(reducing the \(\sqrt{2}\) factor), or by reducing the interval between
observations to reduce the characteristic variability.
\end{abstract}


\section{Introduction}
\label{sec:org8f2c308}
Active Galactic Nuclei (AGN) are among the brightest and most abundant
X-ray sources on the sky. While they are extremely valuable sources to
study in their own right, they can be a nuisance when their emission
is projected onto another object of interest. This is not uncommon in
X-ray observations of clusters of galaxies where the emission from AGN
in, or projected onto, the cluster can be significant relative to the
emission from the cluster. If unresolved, the emission from such AGN
can bias X-ray measurements of the properties of the intra-cluster
medium \citep[e.g.][]{hil10}, or bias the detection of clusters in
X-ray surveys, boosting their detection probability or leading to them
being missed altogether \citep{gil12,som18}.

When high angular resolution X-ray imaging with \emph{Chandra} is
available, any such contaminating AGN can be resolved and excised
efficiently. However, in many applications such as deep observations
of distant clusters (where the greater effective area of \(\XMM\) is
needed), surveys (where the greater grasp - the product of effective
area and field of view - of \(\XMM\) or \emph{eROSITA} is needed), or
observations of cluster outskirts (where \emph{Suzaku}'s lower and more
stable background is needed) \emph{Chandra} is not the optimal primary
instrument. In such cases it is possible to use \emph{Chandra} observations
of the field to detect and characterise AGN so that they may be
masked, subtracted or modelled in other data
\citep[e.g.][]{hil10,tho16}.

The problem with this approach is that the vast majority of AGN show
significant variability in their X-ray flux on timescales from days to
years \citep{pao04}. This introduces an extra source of uncertainty
due to the variability of an AGN between the epoch of the \emph{Chandra}
observation in which it was characterised and the epoch of the
observation in which it must be modelled.

Early work such as \citet{law93,nan97,alm00} investigated the
variability of AGN on timescales of days and weeks finding variability
between about \(10\%\) and \(40\%\). More recent work has extended the
baseline of observations to probe variability on timescales up to 20
years \citep{pao04,mat07,vag11,vag16,mid17}, finding variability
increasing on longer rest-frame timescales, up to about \(50\%\) on
timescales of order 10 years \citep{mid17}.

Where most previous investigations of AGN variability on long timescales have been motivated by the study of the AGN themselves, our motivation and hence our approach is quite different. In particular, we put ourselves in the position of an observer who is faced with a contaminating AGN and we set out to answer the simple question: \emph{given a measurement of the X-ray flux of an AGN at some epoch, but no knowledge of its redshift or nature, what is the uncertainty on its flux at a second epoch due to its intrinsic variability?}

We do this by assembling a sample of AGN that have each been observed
by \emph{Chandra} on two occasions, and then model the pairs of fluxes to
constrain the characteristic variability of the population. Our method
improves on previous analyses by employing a sample selection that
includes non-detections and avoids biasing the inferred variability,
and by using the Poisson statistics of the observed counts to allow
for the inclusion of upper limits in the analysis.

\section{Sample construction and data reduction}
\label{sec:org6a23941}
\label{orgc365af9} In this section we describe the sample definition, data reduction and
analysis used to measure the photon counts in source and background
regions, and hence fluxes for the AGN in our sample.
Our aim was to find pairs of overlapping \emph{Chandra} observations in
order to determine the fluxes of serendipitously observed AGN at two
different epochs. To do this, we considered all public \emph{Chandra}
observations available as of 4th July 2016. We then selected only
ACIS-I observations, and considered pointings whose aim points matched
within \(5\arcmin\) to ensure a reasonable overlap in area. In order to
minimise the occurrence of non-AGN point sources in the data, we
excluded pointings within \(\pm20\degree\) of the Galactic plane and
observations of Galactic targets or nearby galaxies. We then required
observations to have exposures of at least \(20\ks\) to ensure a
reasonable depth. Finally, we defined pairs of matching observations
whose observation time was separated by at least 25 days. This was
chosen so that the separation is at least a factor of 10 larger than
the duration of any of the individual observations.

Where there are a large numbers of observations of a given field, we
selected the longest observations in 25 day blocks and then paired
each with the next available observation separated by at least 25
days, and then repeated until none are left. All pairs are independent
(no observations belong to more than one pair), and if the same AGN is
detected in multiple pairs of observations, only the pair with the
longest interval between observations is used for that AGN.

\begin{figure}
\centering
\includegraphics[angle=0,width=240px]{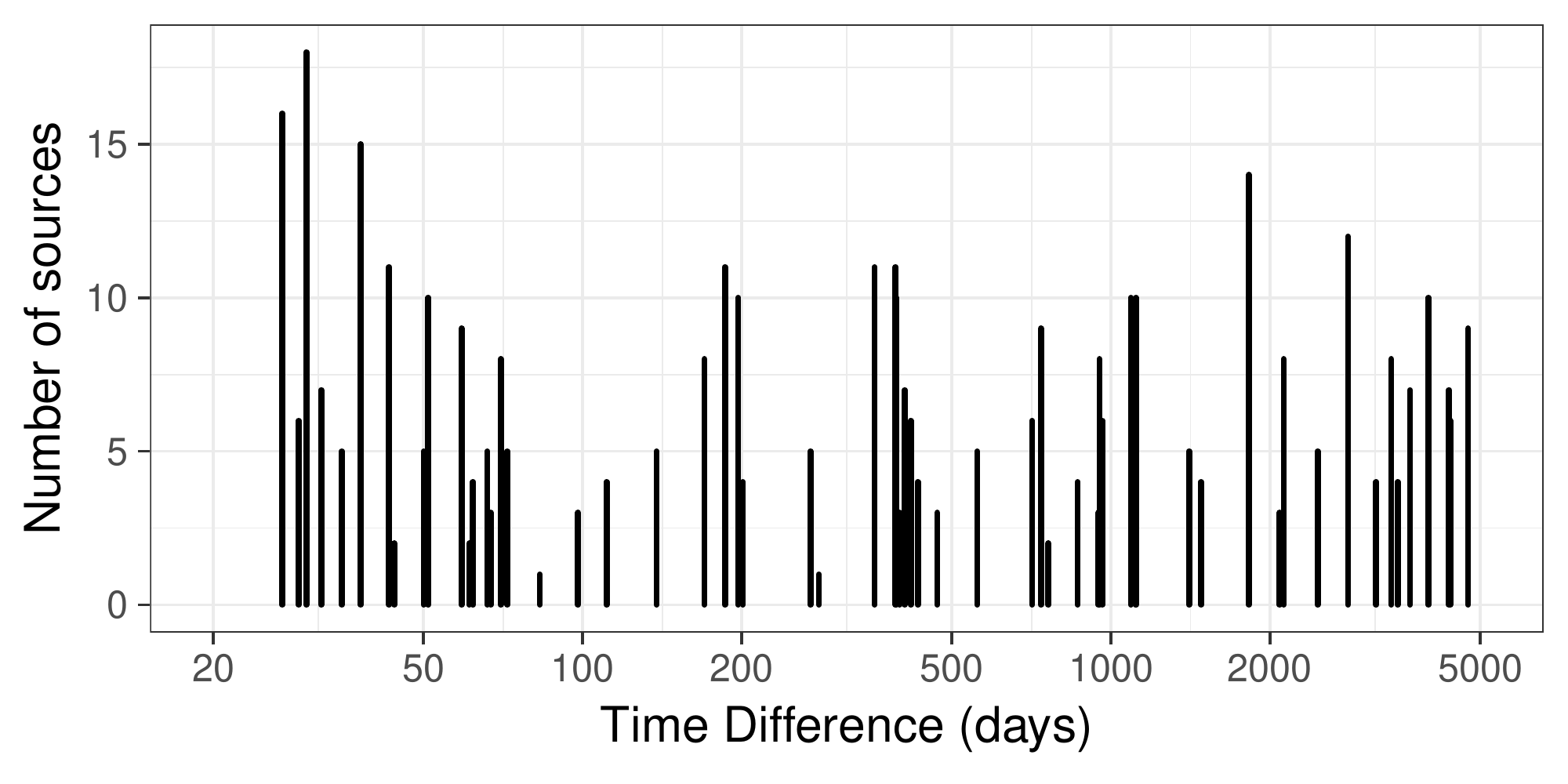}
\caption{\label{fig:org18c59d4}
The separation in time between pairs of observations used to construct the sample}
\end{figure}

This process led to a final dataset of 70 pairs of observations, which
are summarised in Table \ref{tab.obsid}. Fig. \ref{fig:org18c59d4} shows the
time intervals between observations used to construct the sample. The
shortest interval was 27 days and the longest was 4743 days.

\begin{table*}
\centering
\scalebox{0.775}{
\begin{tabular}{lcclccc}
  \hline
  OBSID$_1$ & Exposure$_1$ (ks) & Date$_1$ & OBSID$_2$ & Exposure$_2$ (ks) & Date$_2$ & $\Delta t$ (days) \\
 \hline
  909 & 46.0 & 2000-05-10 & 9371 & 30.7 & 2008-01-18 & 2809 \\
1671 & 166.4 & 2000-11-21 & 3293 & 159.7 & 2001-11-13 & 357 \\
2239 & 130.6 & 2000-12-23 & 8591 & 45.4 & 2007-09-20 & 2462 \\
3185 & 48.0 & 2002-06-14 & 3205 & 30.6 & 2002-10-30 & 138 \\
3197 & 19.9 & 2001-11-12 & 3585 & 15.8 & 2003-01-04 & 418 \\
3280 & 20.3 & 2002-11-03 & 6107 & 15.2 & 2005-11-22 & 1115 \\
3592 & 56.9 & 2003-09-03 & 13999 & 54.4 & 2012-05-14 & 3176 \\
4200 & 59.0 & 2003-01-08 & 1655 & 11.0 & 2001-01-29 & 709 \\
5014 & 32.7 & 2004-08-07 & 3180 & 28.1 & 2003-01-27 & 558 \\
5356 & 96.9 & 2004-08-11 & 3184 & 49.0 & 2002-07-12 & 761 \\
5751 & 128.1 & 2005-06-07 & 513 & 30.6 & 1999-09-22 & 2085 \\
5842 & 46.4 & 2005-03-16 & 6210 & 45.7 & 2005-10-03 & 201 \\
5844 & 45.8 & 2005-03-21 & 6212 & 38.3 & 2005-10-04 & 197 \\
5846 & 49.4 & 2005-03-27 & 6215 & 38.0 & 2005-09-29 & 186 \\
5851 & 35.7 & 2005-10-15 & 6220 & 35.0 & 2005-09-13 & 32 \\
5854 & 50.1 & 2005-09-30 & 6223 & 49.5 & 2005-08-31 & 30 \\
6105 & 37.3 & 2005-06-28 & 3261 & 21.6 & 2002-11-20 & 951 \\
6109 & 37.3 & 2004-12-11 & 9379 & 29.9 & 2008-10-17 & 1406 \\
6110 & 63.2 & 2005-04-20 & 9381 & 29.7 & 2007-12-09 & 963 \\
6217 & 49.5 & 2005-09-23 & 5847 & 44.6 & 2005-04-06 & 170 \\
6930 & 76.1 & 2006-03-06 & 5004 & 19.9 & 2004-02-28 & 737 \\
7998 & 27.6 & 2007-01-10 & 8493 & 19.8 & 2006-12-12 & 29 \\
8122 & 28.8 & 2007-01-20 & 8494 & 20.8 & 2006-12-16 & 35 \\
8471 & 49.4 & 2007-07-29 & 9595 & 27.4 & 2007-09-29 & 62 \\
9425 & 113.5 & 2007-12-24 & 4215 & 18.3 & 2003-12-04 & 1481 \\
9455 & 99.7 & 2008-09-13 & 9729 & 48.1 & 2008-07-09 & 66 \\
9725 & 31.1 & 2008-03-31 & 9450 & 28.8 & 2007-12-11 & 111 \\
9736 & 49.5 & 2008-09-20 & 6219 & 49.5 & 2005-09-25 & 1091 \\
9897 & 69.2 & 2008-08-29 & 13518 & 49.6 & 2011-09-17 & 1114 \\
10769 & 26.7 & 2009-03-20 & 9461 & 23.7 & 2009-06-26 & 98 \\
11710 & 26.7 & 2009-09-09 & 16285 & 19.8 & 2014-09-07 & 1824 \\
11741 & 62.7 & 2009-08-31 & 11870 & 19.8 & 2009-10-20 & 50 \\
11874 & 29.7 & 2010-07-01 & 12092 & 19.8 & 2010-08-08 & 38 \\
11997 & 63.2 & 2010-08-26 & 11742 & 22.5 & 2009-08-29 & 362 \\
12048 & 138.1 & 2010-05-23 & 8595 & 115.4 & 2007-10-19 & 947 \\
12189 & 48.1 & 2011-12-23 & 12180 & 24.7 & 2010-11-20 & 398 \\
12247 & 65.2 & 2010-08-20 & 13138 & 49.4 & 2010-10-10 & 51 \\
12880 & 49.4 & 2010-11-25 & 901 & 38.7 & 1999-12-23 & 3990 \\
12886 & 91.3 & 2010-11-24 & 2204 & 53.9 & 2001-05-05 & 3490 \\
12936 & 34.6 & 2011-01-08 & 11999 & 21.5 & 2009-09-26 & 469 \\
13390 & 38.6 & 2012-06-26 & 925 & 13.8 & 2000-06-22 & 4387 \\
13452 & 72.2 & 2011-09-24 & 13457 & 69.1 & 2011-10-21 & 27 \\
13454 & 91.8 & 2011-09-19 & 13455 & 69.6 & 2011-10-19 & 30 \\
13458 & 116.5 & 2012-11-05 & 3233 & 49.7 & 2002-10-07 & 3682 \\
14022 & 177.4 & 2012-02-21 & 12258 & 59.2 & 2011-01-26 & 391 \\
14333 & 134.8 & 2011-08-31 & 13453 & 69.0 & 2011-10-13 & 43 \\
14407 & 63.2 & 2012-03-16 & 13516 & 39.6 & 2012-12-11 & 270 \\
15173 & 42.5 & 2013-08-14 & 904 & 38.4 & 2000-08-19 & 4743 \\
15658 & 71.6 & 2013-06-23 & 4994 & 13.3 & 2004-03-10 & 3392 \\
16126 & 48.4 & 2014-08-07 & 15123 & 29.3 & 2013-06-26 & 407 \\
16183 & 96.7 & 2014-06-09 & 16456 & 47.5 & 2014-07-29 & 50 \\
16185 & 47.9 & 2016-03-24 & 18730 & 29.7 & 2016-02-02 & 51 \\
16190 & 116.2 & 2014-11-22 & 16178 & 73.9 & 2014-10-07 & 46 \\
16236 & 39.3 & 2014-08-31 & 16237 & 36.5 & 2014-06-09 & 83 \\
16239 & 51.4 & 2015-01-17 & 3589 & 20.0 & 2003-02-07 & 4362 \\
16304 & 97.8 & 2013-11-20 & 16523 & 71.1 & 2014-12-17 & 392 \\
16305 & 93.8 & 2013-12-11 & 16235 & 69.9 & 2013-12-13 & 2 \\
16451 & 112.0 & 2015-03-24 & 17573 & 39.2 & 2015-01-04 & 79 \\
16455 & 89.6 & 2015-10-27 & 18719 & 34.5 & 2015-12-10 & 44 \\
16461 & 111.1 & 2015-05-19 & 16459 & 71.9 & 2015-06-20 & 32 \\
16524 & 44.6 & 2014-05-20 & 12260 & 19.8 & 2012-01-06 & 865 \\
16572 & 44.7 & 2014-02-02 & 9420 & 19.9 & 2008-04-11 & 2123 \\
17296 & 49.3 & 2015-09-07 & 17291 & 49.2 & 2015-10-04 & 27 \\
17299 & 49.3 & 2015-09-10 & 17304 & 44.7 & 2015-07-05 & 67 \\
17303 & 51.2 & 2015-09-18 & 17308 & 44.8 & 2015-07-10 & 70 \\
17306 & 50.8 & 2015-07-08 & 17311 & 48.8 & 2015-09-05 & 59 \\
17307 & 50.8 & 2015-07-09 & 17297 & 49.3 & 2015-09-08 & 61 \\
17599 & 54.4 & 2015-02-15 & 17479 & 49.4 & 2015-04-28 & 72 \\
17628 & 54.8 & 2015-11-18 & 17598 & 51.4 & 2015-02-11 & 280 \\
18822 & 28.7 & 2016-04-14 & 17597 & 23.3 & 2015-02-08 & 431 \\

 \hline
\end{tabular}
}
\caption{Chandra observation pairs used for this analysis. Subscripts 1 and 2 indicate the two observations, where the longer observation is set as observation 1. The listed exposure times are the good times remaining after cleaning. The final column gives the interval in days between the observations.}
\label{tab.obsid}
\end{table*}

As our aim is to determine variability measurements that can be used
to model the flux of an AGN about which nothing other than its flux
and date of observation is known, we make no attempt to cross match
our sample with known AGN. This means that we do not assume knowledge
of the redshift of any sources; we work in terms of the observed flux
and unless noted otherwise, all timescales are in the observer's
frame. This also means that, strictly speaking, we should refer to the
sources we analyse as "X-ray point sources", but given our selection
of fields, the point sources will be dominated by AGN, and we refer to
them as AGN throughout.

\subsection{\emph{Chandra} data analysis}
\label{sec:org7711188}
\label{org87f1dae}
The \(\Chandra\) observations were reduced and analysed using \texttt{CIAO}
version 4.8.2 and \texttt{CALDB} version 4.7.2. The data were reduced
following the standard procedures using the \texttt{chandra\_repro} script,
and the \texttt{deflare} tool to remove periods of high background.

Next, for each pair of observations the astrometry of the
observations was corrected to ensure they matched
closely\footnote{\url{http://cxc.harvard.edu/ciao/threads/reproject_aspect/}}.
This is necessary, as in many cases a source will be detected in one
observation but not in the other, requiring forced photometry at the
source coordinates. If the two observations had a small offset in
astrometry then the aperture would be offset from the source position
in the observation in which the source was not detected. This would
artificially reduce the inferred source flux, biasing the apparent
variability to be higher than the true variability.

To perform the astrometric correction, the longer of the two
observations was defined as observation 1, and the shorter as
observation 2. Sources were detected in each observation in the
\(0.5-7\keV\) energy band using the \texttt{CIAO} \texttt{wavdetect} tool, and those
detected with at least \(7\sigma\) significance were used to register
the images. Observation 2 was corrected to match the astrometry of
observation 1. In all but three pairs of observations, at least 10
sources were available to register the images, with at least 6 sources
used in the other three pairs. The size of the astrometric correction
was typically small. The median correction was \(0.3\arcsec\), and was
less than \(1\arcsec\) in all but three cases (the correction was
smaller than \(1.5\arcsec\) in those cases).

Source detection was then repeated on the corrected observations, and
the source lists for a pair of observations were compared. Sources
whose positions matched within \(1\arcsec\) between the two observations
were considered to be detections of the same source, and we refer to
such a source as a "detected pair" (we demonstrate later that using a
more conservative matching radius of \(0.5\arcsec\) has no impact on our
results). Sources detected in one observation which had no matching
source within \(10\arcsec\) in the other observation were considered to
be undetected in the second observation, and such a source is referred
to as a "detected/undetected pair". Detected sources with a position
match between \(1\arcsec\) and \(10\arcsec\) are excluded as likely
spurious matches.

Next, aperture photometry was performed using the \texttt{CIAO} \texttt{srcflux}
tool in the \(0.5-2\keV\) band. The apertures used for the source
regions were circles with a radius enclosing \(90\%\) of the PSF at
\(1.25\keV\). For the background region, an annulus with an inner radius
equal to the source region and an outer radius five times larger was
used\footnote{For 20 sources, there were no photons detected in either the source or background regions in one of the observations. In these cases the radius of the background region was increased to 15 times that of the source region.}.
In the case of detected pairs, photometry was performed at the
source position determined in each observation. In the case of
detected/undetected pairs, forced photometry was performed in the
observation without a detection at the coordinates of the detected
source. Our requirement that detected/undetected pairs have no other
sources within \(10\arcsec\) ensures that this forced photometry does
not include any contribution from a slightly offset detection of the
same source or other nearby sources.

Fluxes were calculated from the inferred count rates assuming a power
law spectral model with a photon index of 1.7, and were corrected for
Galactic absorption \citep{dic90}. This analysis provided us with
robust flux measurements for all sources, determined using a Bayesian
method to marginalise over the background uncertainty for each source
and takes into account cross-talk between the source and background
regions\footnote{These measurements are performed by the CIAO \texttt{srcflux} tool which uses the algorithm described in
\url{http://cxc.harvard.edu/csc/memos/files/Kashyap_xraysrc.pdf}, which builds on the work of \citet{par06} and references therein.}.
We use the mode of the posterior probability distribution of the flux
as our estimate of the source flux, and in the case where the mode was
zero we define the \(1\sigma\) upper limit on the flux as the value
below which \(68\%\) of the probability density is contained. Note that
in our analysis, these fluxes are only used for selecting subsets of
sources; our likelihood calculations make use of the raw measurements
of the source and background count rates, areas and exposures for each
source.

At this stage we performed some additional filtering of the source
list. Sources falling more than \(6\arcmin\) off-axis in either
observation were rejected to avoid any systematics due to the
increasing PSF (the \(90\%\) encircled energy fraction of the PSF is
\(<5\arcsec\) within this off-axis angle for the energies considered).
This also eliminates cases where a source might be out of the field of
view in one of the two observations. Sources flagged as near chip gaps
by \texttt{srcflux} were excluded. We also rejected 51 sources flagged as
extended by \texttt{wavdetect}. Finally, if a source was detected in multiple
pairs of observations of the same field, only the pair with the
longest interval between observations was retained, leaving a sample
of 1511 unique sources, of which 767 were detected/undetected pairs.

While we use only a subset of these 1511 sources for our variability
analysis, the observed properties of all sources are given in Table
\ref{tab.data} (the table displays the first 60 sources and full table
is available in the electronic version of the paper). Sources were
given a unique identifier of the form \(O_1\_O_2\_i\) where \(O_1\) and \(O_2\)
are the \(\Chandra\) observation identifiers for observation 1 and 2
respectively (where the longest observation is defined to be
observation 1), and \(i\) is an integer indicating the source number in
each pair of observations.

\begin{figure*}
\begin{center}
\scalebox{0.35}{\includegraphics*{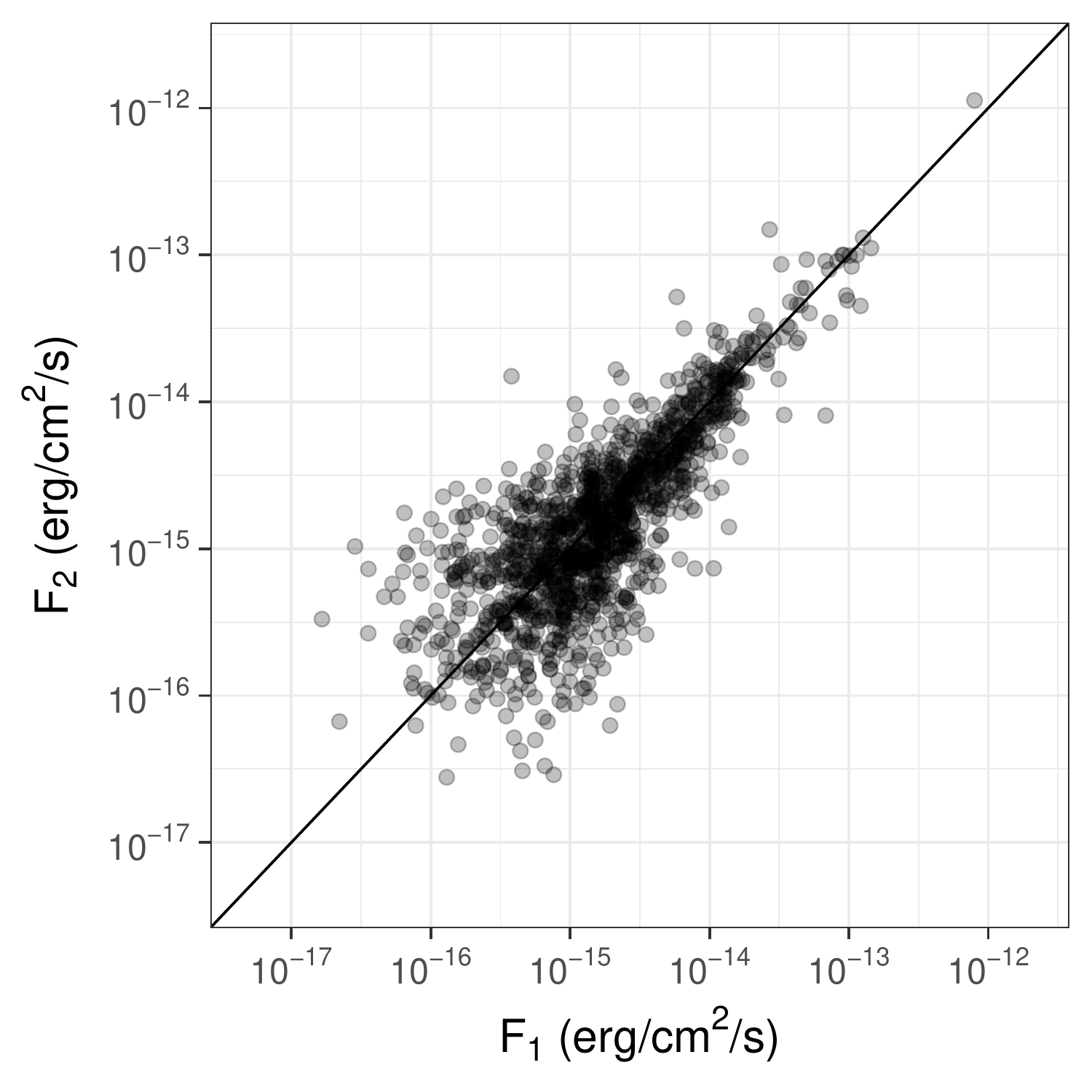}}
\hspace{0.5cm}
\scalebox{0.35}{\includegraphics*{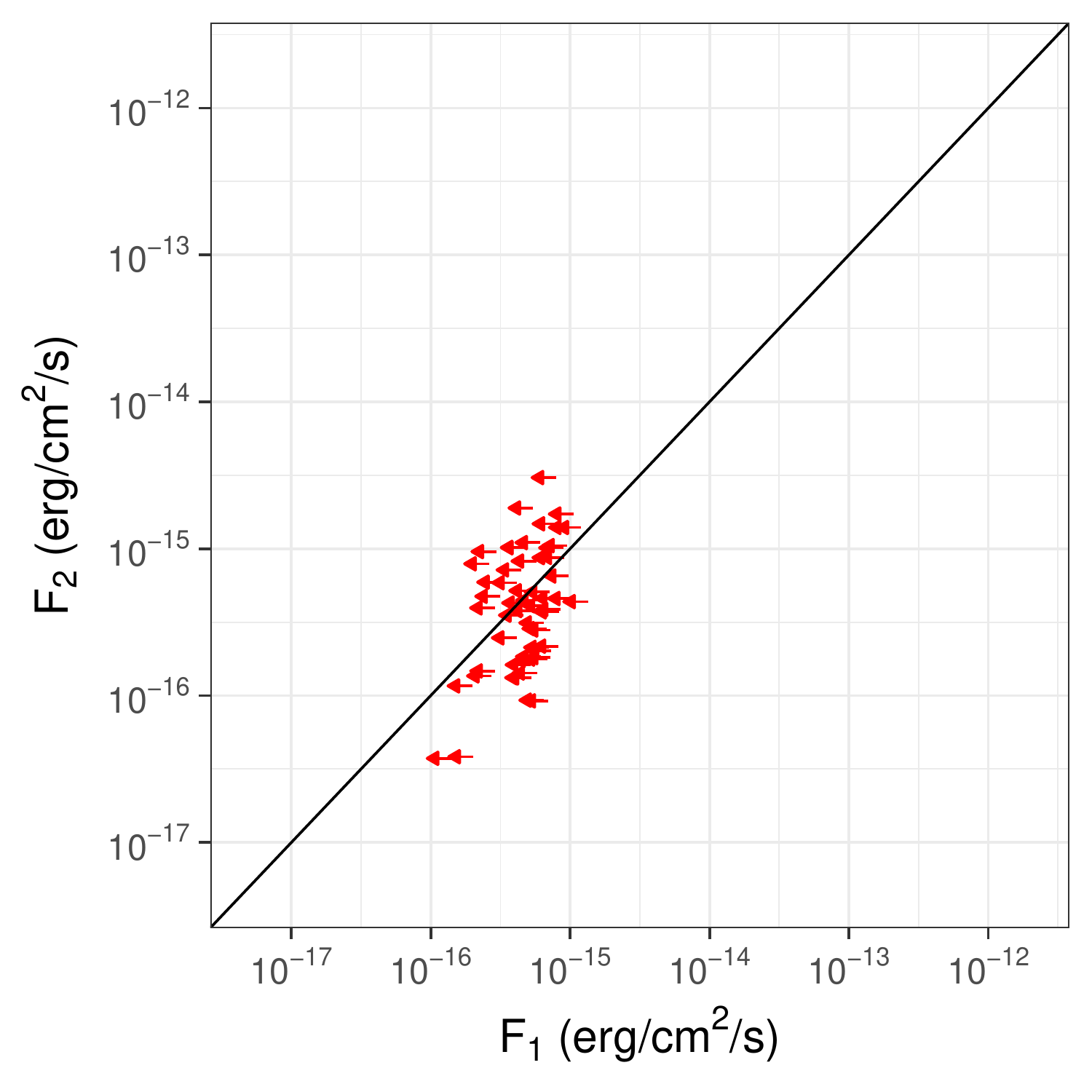}}
\hspace{0.5cm}
\scalebox{0.35}{\includegraphics*{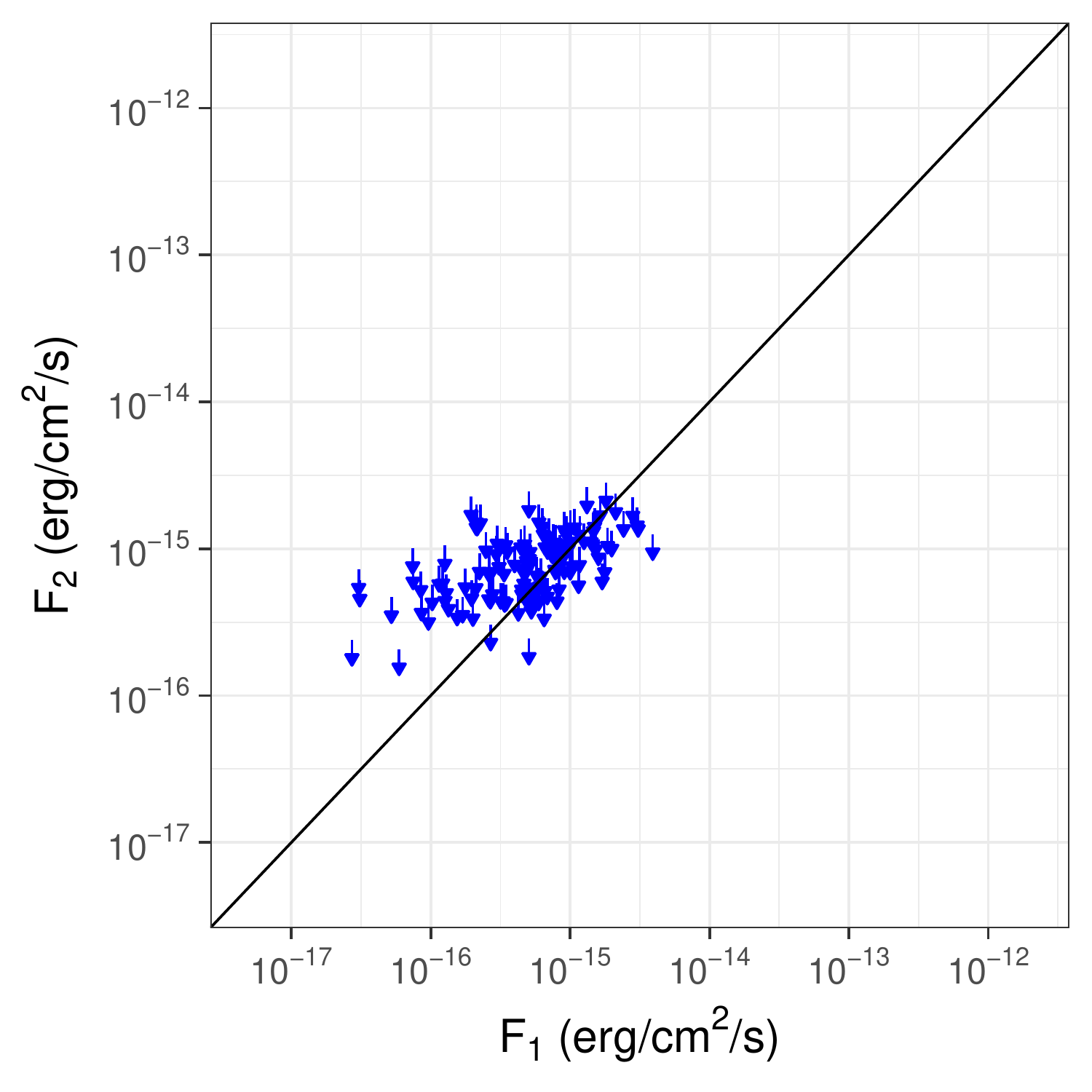}}
\caption[]{\label{fig.f1f2} The fluxes of the AGN in the two observations for the full sample. {\em Left:} AGN with fluxes measured in both observations. {\em Centre:} AGN with upper limits on $F_1$. {\em Right:} AGN with upper limits on $F_2$. The solid line in each panel is a line of equality, and error bars are omitted for clarity. Upper limits are at the $1\sigma$ level.}
\end{center}
\end{figure*}

The fluxes of the AGN in the two observations (\(F_1\) and \(F_2\)
respectively) are plotted in Fig. \ref{fig.f1f2}. The sources with upper limits are
shown in the centre and right panels. There are more upper
limits for \(F_2\) because observation 1 was defined to be the longer of the
two. There are fewer upper limits (205 in total) than the 767
detected/undetected pairs as non-zero fluxes are measured by forced
photometry in many of those cases.

The fluxes scatter about the line of equality as expected, with the
scatter increasing to lower fluxes due to the increased statistical
scatter. The variability in the sources is manifested in the intrinsic
scatter about this line of equality. Measuring this variability relies
on the correct modelling of the statistical uncertainty in the fluxes
(which is non-Gaussian as most sources are in the Poisson regime), and
the inclusion of upper limits. As described in the next section, this
is achieved by expressing the flux variability in terms of the
observed photon counts, rather than using the inferred fluxes
directly.

\section{The AGN variability model}
\label{sec:org450b9e9}
\label{org143b10c}
In this section we will construct a likelihood function relating the
characteristic variability of the AGN in a sample to the observed
source and background counts. This combines the likelihood associated
with the photometric measurements for a pair of fluxes, and the
likelihood of a pair of fluxes given the characteristic variability.
Expressed in this way, the likelihood function properly accounts for
the Poisson nature of the observed counts, which naturally avoids the
need to model upper limits on any inferred fluxes.

\subsection{Photometric measurements}
\label{sec:org5af6b7e}
For a given AGN, the key observed quantities are the photon counts in
the source and background apertures. We denote these as \(T_1,T_2\) in
the source aperture and \(B_1,B_2\) in the background aperture. We wish
to calculate the likelihood of observing these quantities given a set
of model parameters.

We use the terms "source intensity" \(s\) and "background intensity" \(b\)
to refer to the mean number of counts expected in a particular
aperture (i.e. the mean of the Poisson distribution from which the
observed counts are drawn). These are related to the corresponding
fluxes, \(F\) and \(F_b\) respectively, by an energy conversion factor
\(\kappa\), which accounts for the instrument response, source or
background spectral shape, and exposure length. This conversion factor
is defined such that \(F=\kappa s\), and \(\kappa\) is known for each
observation of each AGN.

For a particular observation of a given AGN, the observed counts in
the background aperture \(B\) is a Poisson realisation of the background
intensity \(b\). We can thus write the stochastic relation between the
observed counts and background intensity as
\begin{align}\label{eq.dpoisb}
P(B|b) = \dpois(B|b)
\end{align}
denoting that \(B\) is distributed with a Poisson probability density
with mean \(b\).

For the same AGN, the total counts \(T\) in the source aperture contain a
contribution from the source and background intensities. Thus
\begin{align}\label{eq.dpoist}
P(T|s,b,r) = \dpois(T|s+b/r)
\end{align}
where \(r\) accounts for the difference in detector area \(A\) and
effective area \(E\) between the source and background apertures:
\begin{align}
r = \frac{E_b}{E_s}\frac{A_b}{A_s}
\end{align}

In principle, one can also account for the fact that the PSF scatters
some of the source photons into the background aperture. However,
given that all apertures are defined in the same way to contain 90\% of
the source flux, and that we are interested in variability rather than
absolute flux values, this effect can be neglected (we verified that
including this effect made no significant change to our results).

The background intensity can then be marginalised over to give the joint
likelihood of \(T\) and \(B\) for a given AGN:
\begin{align}
P(T,B|s,r) = \int \dpois(T|s+b/r) \dpois(B|b) \, \diff b
\end{align}
For a pair of observations of the same AGN at different epochs, the
measurements are independent and expressing the intensities in terms
of fluxes (using \(F=\kappa s\)), the joint likelihood is
\begin{multline}\label{eq.TBlik}
P(T_1,B_1,T_2,B_2|F_1,\kappa_1,r_1,F_2,\kappa_2,r_2) = \\
  P(T_1,B_1|F_1,\kappa_1,r_1) \, P(T_2,B_2|F_2,\kappa_2,r_2) \, .
\end{multline}

One of the important features of our model is the inclusion of the
Poisson likelihood to model the observed counts directly, rather than
using derived fluxes and assuming Gaussian statistics. For the main
sample of 416 AGN we define in \textsection \ref{sec.samples}, the
median of the minimum net counts recorded for each AGN in its two
observations is \(\approx22\) (i.e. for half of the sample, the source
has \(\lta22\) net counts in at least one of the two observations).
Assuming Gaussian statistics would therefore be a poor approximation
for a large fraction of the AGN.

\subsection{Characteristic variability}
\label{sec:org9d039a6}
The fundamental aim of this work is to determine the characteristic
amount by which AGN vary in X-ray brightness between observations
separated by months to years. Motivated by this, we model an AGN as
having some mean long-term flux \(F\), with some variability \(\sigma\). We
then assume that fluxes averaged over typical observation lengths (10s
of ks) measured at epochs separated by months to years are sampled
from a lognormal distribution centred on \(\log(F)\) with a standard
deviation \(\sigma\). Working in natural log space, \(\sigma\) then
represents the characteristic fractional variability of the AGN on the
timescales probed. We further assume that the variability of all AGN
can be described by similar lognormal distributions, each with a
different mean flux, but all sharing the same fractional variability
\(\sigma\). Our aim is then to determine the value of \(\sigma\), the
characteristic fractional X-ray variability of AGN between observation
epochs. This assumption of log-normality is similar to many previous
studies of variability in ensembles of AGN, which have measured the
average fractional variability \citep[e.g.][]{alm00,mat07}

In this model, the probability distribution for a pair
of fluxes \(F_1,F_2\) is
\begin{multline} \label{eq.pF1F2}
P(F_1,F_2|F,\sigma) = \\
 \dlnorm(F_1|\log(F),\sigma)\, \dlnorm(F_2 |\log(F),\sigma) \, ,
\end{multline}
where dlnorm is the lognormal probability density function, and \(\sigma\)
is the quantity in which we are interested, describing the fractional
variability of the AGN.

It is known a-priori that AGN fluxes are not uniformly distributed,
but instead follow a distribution (generically referred to as a
\(\lognlogs\) distribution) which can be approximated as a power-law or
broken power-law \citep[e.g.][]{has98,mat08,leh12}. We therefore write
the prior probability density on \(F\) as
\begin{align}\label{eq.powpdf}
P(F|\beta) = C (F)^{-\beta},
\end{align}
i.e. the source flux (\(F\)) is distributed with a power-law probability
density with a negative slope \(\beta\), and \(C\) is a normalisation
factor computed to normalise the density to unity over the range of
flux considered. The slope of the \(\lognlogs\) distribution \(\beta\) is
a nuisance parameter in our model, but as we will see later, with a
suitable sample definition our constraints on the variability are
insensitive to this parameter.

The mean flux can then be marginalised over, and combining Equations
Eq. \ref{eq.pF1F2} and \ref{eq.powpdf} we then obtain
\begin{align} \label{eq.pF1F2sig}
P(F_1,F_2|\sigma,\beta) = \int P(F_1,F_2|F,\sigma) P(F|\beta) \, \diff F \, .
\end{align}

\clearpage
\begin{turnpage}
\begin{table}
\centering
\scalebox{0.55}{
\begin{tabular}{lcccccccccccccccccccccccc}
  \hline
  ID & Det$_1$ & Det$_2$ & RA$_1$ & Dec$_1$ & $T_1$ & $A_{s1}$ &
                                                                 $E_{s1}/10^{6}$ & $B_1$ & $A_{b1}$ & $E_{b1}/10^{6}$ & $\kappa_{1}/10^{-15}$ & $F_{1}/10^{-15}$ & RA$_2$ & Dec$_2$ & $T_2$ & $A_{s2}$ & $E_{s2}/10^{6}$ & $B_2$ & $A_{b2}/10^{6}$ & $E_{b2}$ & $\kappa_{2}/10^{-15}$ & $F_{2}/10^{-15}$ & Separation & $\Delta t$\\
  &  &  & (J2000) & (J2000) & counts & pixels & cm$^2$\,s & counts & pixels & cm$^2$\,s & $\flux$ & $\flux$ & (J2000) & (J2000) & counts & pixels & cm$^2$\,s & counts & pixels & cm$^2$\,s & $\flux$ & $\flux$ & arcsec & days \\
 \hline
  $11997\_11742\_7$ & 1 & 0 & 0.0658 & -50.1777 & 6 & 70.1 & 27.1 & 16 & 1692.3 & 23.3 & 0.12 & 0.69 & 0.0658 & -50.1777 & 0 & 70.1 & 9.2 & 7 & 1690.4 & 8.6 & 0.33 & 0.85$^\dagger$ & 0.0 & 362 \\
  $11997\_11742\_6$ & 1 & 0 & 0.0960 & -50.1940 & 12 & 186.9 & 26.6 & 51 & 4503.9 & 26.4 & 0.12 & 1.31 & 0.0960 & -50.1940 & 5 & 186.7 & 9.7 & 11 & 4499.0 & 9.6 & 0.32 & 1.63 & 0.0 & 362 \\
  $6105\_3261\_52$ & 1 & 0 & 2.8468 & -15.4007 & 2 & 144.3 & 15.5 & 10 & 3407.5 & 15.1 & 0.20 & 0.35 & 2.8469 & -15.4007 & 0 & 287.8 & 9.4 & 7 & 6588.8 & 9.6 & 0.33 & 0.85$^\dagger$ & 0.0 & 951 \\
  $6105\_3261\_3$ & 1 & 1 & 2.8820 & -15.3890 & 231 & 29.6 & 15.7 & 14 & 712.0 & 14.9 & 0.20 & 52.2 & 2.8820 & -15.3890 & 120 & 100.8 & 10.3 & 11 & 2328.9 & 10.1 & 0.30 & 40.2 & 0.2 & 951 \\
  $6105\_3261\_11$ & 1 & 1 & 2.8829 & -15.4318 & 30 & 134.6 & 16.3 & 10 & 3238.0 & 16.3 & 0.19 & 6.4 & 2.8829 & -15.4318 & 9 & 69.3 & 9.6 & 3 & 1674.6 & 9.5 & 0.32 & 3.13 & 0.4 & 951 \\
  $6105\_3261\_66$ & 0 & 1 & 2.8914 & -15.4547 & 2 & 272.9 & 16.0 & 17 & 6561.2 & 16.0 & 0.20 & 0.28 & 2.8914 & -15.4547 & 5 & 68.8 & 10.3 & 3 & 1655.0 & 10.2 & 0.30 & 1.61 & 0.0 & 951 \\
  $6105\_3261\_41$ & 1 & 0 & 2.9066 & -15.3891 & 3 & 13.5 & 14.2 & 9 & 331.1 & 15.2 & 0.22 & 0.67 & 2.9066 & -15.3891 & 4 & 38.4 & 10.4 & 17 & 927.4 & 10.4 & 0.30 & 1.1 & 0.0 & 951 \\
  $6105\_3261\_2$ & 1 & 1 & 2.9163 & -15.3946 & 207 & 14.9 & 17.3 & 47 & 361.2 & 17.3 & 0.18 & 42 & 2.9163 & -15.3946 & 76 & 23.1 & 10.5 & 22 & 556.6 & 10.5 & 0.29 & 25.2 & 0.2 & 951 \\
  $6105\_3261\_37$ & 1 & 0 & 2.9172 & -15.4291 & 8 & 84.7 & 16.8 & 4 & 2040.4 & 16.8 & 0.19 & 1.63 & 2.9172 & -15.4291 & 1 & 13.4 & 10.5 & 1 & 327.6 & 10.5 & 0.29 & 0.3 & 0.0 & 951 \\
  $6105\_3261\_72$ & 0 & 1 & 2.9179 & -15.4341 & 2 & 104.2 & 16.8 & 8 & 2501.6 & 16.7 & 0.19 & 0.34 & 2.9179 & -15.4341 & 3 & 13.5 & 10.5 & 1 & 338.5 & 10.5 & 0.29 & 0.94 & 0.0 & 951 \\
  $6105\_3261\_44$ & 1 & 0 & 2.9186 & -15.3696 & 6 & 8.8 & 16.7 & 4 & 213.4 & 16.6 & 0.19 & 1.22 & 2.9186 & -15.3696 & 2 & 60.8 & 10.2 & 28 & 1467.6 & 10.2 & 0.31 & 0.29 & 0.0 & 951 \\
  $6105\_3261\_54$ & 1 & 0 & 2.9204 & -15.3499 & 2 & 9.5 & 16.5 & 0 & 233.8 & 15.9 & 0.19 & 0.37 & 2.9204 & -15.3500 & 1 & 126.5 & 10.1 & 11 & 3051.6 & 9.9 & 0.31 & 0.18 & 0.0 & 951 \\
  $6105\_3261\_35$ & 1 & 0 & 2.9205 & -15.4006 & 12 & 20.1 & 17.3 & 21 & 481.9 & 17.2 & 0.18 & 2.25 & 2.9206 & -15.4006 & 3 & 15.6 & 10.5 & 8 & 380.5 & 10.5 & 0.29 & 0.89 & 0.0 & 951 \\
  $6105\_3261\_30$ & 1 & 1 & 2.9253 & -15.3680 & 2 & 8.9 & 17.4 & 2 & 215.3 & 17.4 & 0.18 & 0.38 & 2.9253 & -15.3681 & 42 & 57.2 & 9.2 & 3 & 1375.1 & 9.0 & 0.33 & 14.9 & 0.5 & 951 \\
  $6105\_3261\_12$ & 1 & 1 & 2.9286 & -15.4309 & 29 & 92.6 & 15.2 & 11 & 2224.4 & 15.7 & 0.22 & 6.91 & 2.9286 & -15.4309 & 14 & 10.1 & 10.5 & 3 & 248.2 & 10.5 & 0.29 & 4.44 & 0.2 & 951 \\
  $6105\_3261\_67$ & 0 & 1 & 2.9409 & -15.4454 & 2 & 168.6 & 16.6 & 17 & 4054.9 & 16.6 & 0.19 & 0.27 & 2.9409 & -15.4454 & 0 & 11.7 & 9.9 & 5 & 2659.5 & 9.2 & 0.30 & 0.77$^\dagger$ & 0.0 & 951 \\
  $6105\_3261\_4$ & 1 & 1 & 2.9475 & -15.3888 & 61 & 18.5 & 17.5 & 14 & 447.8 & 17.5 & 0.18 & 12.2 & 2.9475 & -15.3889 & 8 & 15.4 & 10.4 & 8 & 379.6 & 10.4 & 0.30 & 2.6 & 0.2 & 951 \\
  $6105\_3261\_6$ & 1 & 1 & 2.9491 & -15.3742 & 72 & 13.2 & 17.4 & 10 & 319.9 & 17.4 & 0.18 & 14.9 & 2.9491 & -15.3743 & 24 & 32.4 & 8.9 & 3 & 790.0 & 9.1 & 0.36 & 9.33 & 0.1 & 951 \\
  $6105\_3261\_65$ & 0 & 1 & 2.9498 & -15.3784 & 5 & 14.5 & 17.5 & 8 & 349.5 & 17.5 & 0.18 & 0.94 & 2.9498 & -15.3784 & 4 & 27.2 & 9.5 & 6 & 658.1 & 9.5 & 0.32 & 1.35 & 0.0 & 951 \\
  $6105\_3261\_1$ & 1 & 1 & 2.9506 & -15.4154 & 40 & 61.7 & 17.2 & 11 & 1487.4 & 17.2 & 0.18 & 8.12 & 2.9506 & -15.4154 & 12 & 8.8 & 10.4 & 7 & 213.6 & 10.4 & 0.29 & 3.87 & 0.3 & 951 \\
  $6105\_3261\_51$ & 1 & 0 & 2.9538 & -15.4039 & 3 & 39.6 & 17.3 & 11 & 955.4 & 17.3 & 0.18 & 0.51 & 2.9538 & -15.4039 & 0 & 9.9 & 10.4 & 2 & 241.4 & 10.3 & 0.29 & 0.76$^\dagger$ & 0.0 & 951 \\
  $6105\_3261\_7$ & 1 & 1 & 2.9711 & -15.3649 & 25 & 32.1 & 16.1 & 6 & 775.2 & 14.4 & 0.19 & 5.41 & 2.9711 & -15.3649 & 12 & 68.8 & 10.0 & 7 & 1654.0 & 10.0 & 0.31 & 4.06 & 0.1 & 951 \\
  $6105\_3261\_62$ & 0 & 1 & 2.9735 & -15.4447 & 1 & 267.1 & 16.4 & 18 & 6418.3 & 16.4 & 0.19 & 0.05 & 2.9735 & -15.4447 & 2 & 17.5 & 10.4 & 0 & 427.2 & 10.4 & 0.29 & 0.58 & 0.0 & 951 \\
  $15173\_904\_17$ & 1 & 1 & 10.3600 & -9.3700 & 53 & 85.6 & 14.9 & 67 & 2070.4 & 14.7 & 0.23 & 12.9 & 10.3600 & -9.3700 & 76 & 146.9 & 17.2 & 132 & 3547.4 & 17.1 & 0.17 & 13.9 & 0.3 & 4743 \\
  $15173\_904\_36$ & 1 & 0 & 10.3739 & -9.4157 & 4 & 199.7 & 14.8 & 67 & 4796.5 & 14.6 & 0.23 & 0.31 & 10.3739 & -9.4157 & 4 & 140.6 & 18.4 & 78 & 3373.9 & 18.0 & 0.17 & 0.14 & 0.0 & 4743 \\
  $15173\_904\_45$ & 0 & 1 & 10.3952 & -9.4342 & 2 & 279.4 & 16.0 & 94 & 6708.9 & 15.8 & 0.21 & 0.80$^\dagger$ & 10.3952 & -9.4342 & 10 & 125.9 & 18.7 & 58 & 3035.3 & 18.4 & 0.16 & 1.39 & 0.0 & 4743 \\
  $15173\_904\_7$ & 1 & 1 & 10.3955 & -9.3358 & 15 & 12.9 & 15.9 & 18 & 310.5 & 16.3 & 0.21 & 3.29 & 10.3954 & -9.3357 & 19 & 62.8 & 18.6 & 97 & 1519.6 & 18.2 & 0.16 & 2.68 & 0.2 & 4743 \\
  $15173\_904\_20$ & 1 & 1 & 10.4020 & -9.3232 & 7 & 13.2 & 15.9 & 23 & 320.0 & 15.6 & 0.21 & 1.44 & 10.4020 & -9.3231 & 16 & 77.0 & 18.9 & 143 & 1854.1 & 18.5 & 0.16 & 1.76 & 0.3 & 4743 \\
  $15173\_904\_1$ & 1 & 1 & 10.4021 & -9.3943 & 27 & 53.0 & 17.0 & 32 & 1278.2 & 17.0 & 0.20 & 5.77 & 10.4021 & -9.3943 & 40 & 28.7 & 19.1 & 22 & 688.5 & 18.0 & 0.16 & 6.95 & 0.7 & 4743 \\
  $15173\_904\_2$ & 1 & 1 & 10.4159 & -9.3883 & 23 & 32.4 & 17.2 & 34 & 778.1 & 17.2 & 0.20 & 4.82 & 10.4159 & -9.3883 & 32 & 14.1 & 18.7 & 20 & 343.5 & 18.6 & 0.16 & 5.91 & 0.3 & 4743 \\
  $15173\_904\_5$ & 1 & 1 & 10.4195 & -9.3556 & 53 & 9.1 & 17.5 & 28 & 224.1 & 17.5 & 0.19 & 11.4 & 10.4195 & -9.3556 & 73 & 15.2 & 19.6 & 47 & 370.8 & 19.4 & 0.15 & 12.3 & 0.3 & 4743 \\
  $15173\_904\_48$ & 0 & 1 & 10.4584 & -9.3533 & 5 & 20.6 & 17.6 & 72 & 501.4 & 17.6 & 0.19 & 0.45 & 10.4584 & -9.3533 & 11 & 12.7 & 17.5 & 41 & 305.1 & 16.8 & 0.17 & 1.83 & 0.0 & 4743 \\
  $15173\_904\_55$ & 0 & 1 & 10.4697 & -9.3364 & 7 & 31.5 & 16.5 & 146 & 756.9 & 16.5 & 0.21 & 0.21 & 10.4697 & -9.3364 & 19 & 29.6 & 19.3 & 214 & 712.9 & 19.3 & 0.16 & 1.79 & 0.0 & 4743 \\
  $15173\_904\_11$ & 1 & 1 & 10.4718 & -9.4185 & 29 & 229.5 & 14.7 & 152 & 5513.4 & 14.9 & 0.23 & 5.96 & 10.4718 & -9.4186 & 27 & 40.8 & 15.9 & 29 & 992.5 & 17.3 & 0.19 & 5.55 & 0.6 & 4743 \\
  $15173\_904\_15$ & 1 & 1 & 10.4748 & -9.3792 & 17 & 78.9 & 16.2 & 126 & 1895.9 & 16.2 & 0.21 & 2.78 & 10.4748 & -9.3792 & 19 & 14.5 & 19.8 & 35 & 348.9 & 19.8 & 0.15 & 3.05 & 0.3 & 4743 \\
  $15173\_904\_13$ & 1 & 1 & 10.4893 & -9.4113 & 87 & 273.6 & 16.4 & 151 & 6517.5 & 15.9 & 0.21 & 18.9 & 10.4893 & -9.4113 & 121 & 57.6 & 19.5 & 46 & 1389.2 & 19.0 & 0.16 & 20.9 & 0.1 & 4743 \\
  $15173\_904\_49$ & 0 & 1 & 10.5063 & -9.3335 & 21 & 156.0 & 14.3 & 359 & 3750.4 & 15.1 & 0.24 & 1.61 & 10.5063 & -9.3335 & 39 & 109.0 & 15.1 & 283 & 2615.1 & 17.1 & 0.20 & 6.17 & 0.0 & 4743 \\
  $15173\_904\_43$ & 1 & 0 & 10.5184 & -9.3308 & 29 & 254.1 & 15.2 & 406 & 6118.1 & 15.4 & 0.23 & 3.07 & 10.5185 & -9.3308 & 11 & 167.9 & 15.4 & 334 & 4036.9 & 16.8 & 0.19 & 1.21$^\dagger$ & 0.0 & 4743 \\
  $13458\_3233\_62$ & 1 & 0 & 13.8713 & 26.4314 & 16 & 203.9 & 42.3 & 41 & 4894.6 & 40.7 & 0.08 & 1.34 & 13.8712 & 26.4314 & 1 & 72.6 & 23.6 & 4 & 1743.6 & 23.6 & 0.14 & 0.12 & 0.0 & 3682 \\
  $13458\_3233\_15$ & 1 & 1 & 13.8797 & 26.4147 & 129 & 111.8 & 45.9 & 26 & 2712.6 & 44.3 & 0.08 & 11.1 & 13.8797 & 26.4147 & 78 & 44.3 & 22.2 & 8 & 1068.6 & 21.1 & 0.14 & 12.5 & 0.2 & 3682 \\
  $13458\_3233\_61$ & 1 & 0 & 13.8843 & 26.3699 & 5 & 69.3 & 42.7 & 12 & 1674.6 & 43.6 & 0.08 & 0.41 & 13.8843 & 26.3699 & 6 & 80.3 & 20.3 & 11 & 1930.1 & 19.9 & 0.16 & 0.98 & 0.0 & 3682 \\
  $13458\_3233\_75$ & 0 & 1 & 13.8903 & 26.4538 & 2 & 232.2 & 44.5 & 66 & 5366.0 & 42.6 & 0.08 & 0.32$^\dagger$ & 13.8902 & 26.4538 & 5 & 59.8 & 16.6 & 10 & 1437.6 & 19.6 & 0.20 & 1.01 & 0.0 & 3682 \\
  $13458\_3233\_26$ & 1 & 1 & 13.8936 & 26.4553 & 20 & 228.3 & 45.7 & 70 & 5246.9 & 44.6 & 0.08 & 1.48 & 13.8935 & 26.4553 & 5 & 57.1 & 20.5 & 5 & 1364.7 & 19.3 & 0.16 & 0.84 & 0.3 & 3682 \\
  $13458\_3233\_23$ & 1 & 1 & 13.8964 & 26.4329 & 7 & 109.8 & 43.6 & 31 & 2639.4 & 43.2 & 0.08 & 0.52 & 13.8964 & 26.4329 & 3 & 27.1 & 24.1 & 2 & 653.4 & 23.8 & 0.13 & 0.43 & 0.7 & 3682 \\
  $13458\_3233\_70$ & 0 & 1 & 13.9046 & 26.4086 & 0 & 41.5 & 46.6 & 20 & 998.5 & 45.4 & 0.07 & 0.19$^\dagger$ & 13.9046 & 26.4086 & 1 & 16.4 & 22.5 & 1 & 399.6 & 22.2 & 0.14 & 0.15 & 0.0 & 3682 \\
  $13458\_3233\_4$ & 1 & 1 & 13.9049 & 26.3563 & 35 & 34.6 & 43.4 & 24 & 836.0 & 43.8 & 0.08 & 3.03 & 13.9049 & 26.3562 & 20 & 77.5 & 22.4 & 12 & 1863.1 & 21.9 & 0.14 & 3.08 & 0.4 & 3682 \\
  $13458\_3233\_59$ & 1 & 0 & 13.9130 & 26.4648 & 11 & 226.8 & 41.1 & 74 & 5454.7 & 40.7 & 0.09 & 0.78 & 13.9130 & 26.4648 & 2 & 51.2 & 23.6 & 10 & 1232.0 & 23.6 & 0.14 & 0.24 & 0.0 & 3682 \\
  $13458\_3233\_11$ & 1 & 1 & 13.9151 & 26.3972 & 11 & 20.9 & 41.4 & 10 & 508.5 & 39.7 & 0.08 & 0.98 & 13.9151 & 26.3972 & 6 & 13.5 & 24.0 & 7 & 340.8 & 24.0 & 0.13 & 0.88 & 0.2 & 3682 \\
  $13458\_3233\_1$ & 1 & 1 & 13.9202 & 26.3410 & 24 & 33.8 & 44.5 & 9 & 815.0 & 45.9 & 0.08 & 2.02 & 13.9202 & 26.3410 & 12 & 113.6 & 21.1 & 22 & 2732.8 & 21.5 & 0.15 & 1.84 & 0.4 & 3682 \\
  $13458\_3233\_50$ & 1 & 0 & 13.9396 & 26.3687 & 8 & 10.1 & 48.9 & 9 & 244.6 & 48.9 & 0.07 & 0.6 & 13.9396 & 26.3687 & 2 & 27.3 & 19.4 & 7 & 677.5 & 20.2 & 0.17 & 0.32 & 0.0 & 3682 \\
  $13458\_3233\_3$ & 1 & 1 & 13.9402 & 26.3544 & 31 & 14.4 & 48.7 & 15 & 346.5 & 48.7 & 0.07 & 2.47 & 13.9402 & 26.3543 & 23 & 58.4 & 22.8 & 17 & 1405.5 & 22.8 & 0.14 & 3.59 & 0.4 & 3682 \\
  $13458\_3233\_12$ & 1 & 1 & 13.9427 & 26.3988 & 34 & 10.2 & 49.3 & 9 & 249.0 & 48.8 & 0.07 & 2.62 & 13.9427 & 26.3988 & 15 & 9.1 & 24.0 & 9 & 228.4 & 24.0 & 0.14 & 2.29 & 0.2 & 3682 \\
  $13458\_3233\_29$ & 1 & 1 & 13.9448 & 26.4634 & 12 & 160.4 & 43.1 & 64 & 3846.6 & 43.8 & 0.08 & 0.88 & 13.9448 & 26.4634 & 12 & 34.9 & 23.9 & 11 & 848.8 & 23.3 & 0.14 & 1.79 & 0.7 & 3682 \\
  $13458\_3233\_10$ & 1 & 1 & 13.9588 & 26.3948 & 27 & 9.1 & 48.9 & 10 & 225.7 & 48.7 & 0.07 & 2.17 & 13.9588 & 26.3948 & 20 & 11.3 & 24.0 & 17 & 275.3 & 24.0 & 0.14 & 3.12 & 0.1 & 3682 \\
  $13458\_3233\_44$ & 1 & 1 & 13.9597 & 26.4698 & 15 & 209.4 & 44.2 & 59 & 5031.1 & 44.3 & 0.08 & 1.13 & 13.9597 & 26.4698 & 3 & 58.1 & 23.8 & 7 & 1397.1 & 23.1 & 0.14 & 0.41 & 0.3 & 3682 \\
  $13458\_3233\_51$ & 1 & 0 & 13.9654 & 26.4275 & 38 & 33.2 & 48.8 & 194 & 799.7 & 48.7 & 0.07 & 2.44 & 13.9654 & 26.4275 & 3 & 11.9 & 24.1 & 39 & 290.7 & 24.1 & 0.14 & 0.21 & 0.0 & 3682 \\
  $13458\_3233\_54$ & 1 & 0 & 13.9662 & 26.3230 & 33 & 54.9 & 42.9 & 106 & 1320.2 & 44.8 & 0.08 & 2.58 & 13.9662 & 26.3229 & 19 & 241.8 & 19.6 & 218 & 5797.9 & 20.6 & 0.17 & 1.87 & 0.0 & 3682 \\
  $13458\_3233\_73$ & 0 & 1 & 13.9718 & 26.4691 & 7 & 215.8 & 42.4 & 61 & 5187.9 & 44.1 & 0.08 & 0.41 & 13.9718 & 26.4692 & 3 & 68.9 & 23.8 & 14 & 1661.4 & 23.0 & 0.14 & 0.37 & 0.0 & 3682 \\
  $13458\_3233\_55$ & 1 & 0 & 13.9860 & 26.3771 & 22 & 13.4 & 48.4 & 32 & 327.6 & 48.4 & 0.07 & 1.72 & 13.9860 & 26.3771 & 7 & 47.8 & 23.4 & 35 & 1155.1 & 23.4 & 0.14 & 0.89 & 0.0 & 3682 \\
  $13458\_3233\_9$ & 1 & 1 & 13.9866 & 26.3903 & 8 & 14.5 & 48.5 & 39 & 349.7 & 48.4 & 0.07 & 0.52 & 13.9866 & 26.3903 & 4 & 32.1 & 23.8 & 42 & 775.7 & 23.8 & 0.14 & 0.35 & 1.0 & 3682 \\
   \hline
\end{tabular}
}
\caption{Properties of the detected sources. ID is a unique identifier of each source of the form $O1\_O2\_I$ where $O1$ and $O2$ are the $\Chandra$ observation identifiers for observation 1 and 2 respectively, and $I$ is an integer indicating the source number in each pair of observations. Observation 1 is defined to be the longer of the two observations. For the other columns, subscripts 1 and 2 indicate whether the quantity is measured for observation 1 or 2, and subscripts $s$ and $b$ indicate if the quantity is measured for the source or background region, respectively. "Det" is a binary flag indicating whether a source was detected in a given observation (if not then forced photometry was performed at the source position determined from the other observation). RA and Dec give the source coordinates in degrees. $T$ gives the total number of counts in the source region, $A$ gives the area of each region in pixels ($0.492\arcsec$ on a side), $E$ gives the mean exposure map value in each region, $B$ gives the total counts in the background region, and $\kappa$ gives the energy conversion factor for the source. $F$ is the unabsorbed flux of the source, determined from the mode of the posterior probability distribution of the flux, where values marked with $\dagger$ are $1\sigma$ upper limits. The "Separation" column gives the angular separation between detections of the same source in the two observations, and $\Delta t$ gives the separation in days between the two observations. All photometric quanities are measured in the $0.5-2\keV$ band in the observer's frame. This is a truncated version of the table; the full data for all 1511 sources are presented in the online version of the paper.}
\label{tab.data}
\end{table}

\clearpage
\end{turnpage}

\subsection{The final likelihood}
\label{sec:org65dfe3a}
The likelihood of the counts observed in a pair of observations of an
AGN, given \(\sigma\) and \(\beta\) can now be written by combining
equations \ref{eq.TBlik} and \ref{eq.pF1F2sig} and marginalising out
\(F_1\) and \(F_2\) to give
\begin{multline}\label{eq.likelihood}
P(T_1,B_1,T_2,B_2|\sigma,\beta,\kappa_1,r_1,\kappa_2,r_2) = \\
  \int \int P(T_1,B_1,T_2,B_2|F_1,\kappa_1,r_1,F_2,\kappa_2,r_2) \times \\
   P(F_1,F_2|\sigma,\beta) \,\diff F_1 \,\diff F_2
\end{multline}

The final likelihood for a sample of \(N\) AGN is the product of the
individual probabilities:
\begin{align}
\lik = \prod_{i=1}^N P(T_{1,i},B_{1,i},T_{2,i},B_{2,i}|\sigma,\beta,\kappa_{1,i},r_{1,i},\kappa_{2,i},r_{2,i})
\end{align}

In principle one could multiply this likelihood by priors on \(\sigma\)
and \(\beta\) and treat it as a posterior distribution to be sampled
with standard Bayesian techniques. However, evaluating this likelihood
function for a sample size of a few hundred AGN is computationally
expensive due primarily to the Poisson probability evaluations inside
nested integrals. This can be mitigated by splitting the sources over
multiple CPU cores to evaluate their likelihoods in parallel and
combining these for final likelihood. Even so, for a typical fit of
\(\sim300\) AGN spread across 30 CPU cores (using more cores leads to
diminishing returns due to overheads), each likelihood evaluation took
around \(20\s\) of wall time. For this reason we adopted a simple
maximum-likelihood analysis, and given the model has just two
parameters (or one when \(\beta\) is fixed), straightforward grid searches
were sufficient to map the likelihood distribution.

Parameter uncertainties were estimated using the likelihood ratio
method \citep[e.g.][]{cas79}, noting that \(2\log{\lik}\) is \(\chisq\)
distributed with a number of degrees of freedom equal to the number of
model parameters. Thus for a two parameter fit the \((1\sigma, 2\sigma,
3\sigma)\) confidence intervals enclose the parameter values for which
\(2\log{\lik_0}-2\log{\lik}>(2.3,6.0,11.6)\), where \(\lik_0\) is the
maximum value of the likelihood. For a single parameter fit, the
corresponding levels are \(2\log{\lik_0}-2\log{\lik}>(0.98,3.8,8.8)\).

\subsection{Subsample definition and selection biases}
\label{sec:orgc1a3037}
\label{sec.samples} The selection of the sample used for this type of
ensemble variability analysis can easily result in biases that will
increase or decrease the apparent value of \(\sigma\). With no
additional selection applied, our sample of AGN is defined by the
requirement that each AGN be detected by \texttt{wavdetect} in at least one
of its two observations. In principle this selection function could be
modelled with simulations but it is possible instead to define a
subsample for which the selection results in an unbiased estimate of
\(\sigma\).

\begin{figure*}
\begin{center}
\scalebox{0.35}{\includegraphics*{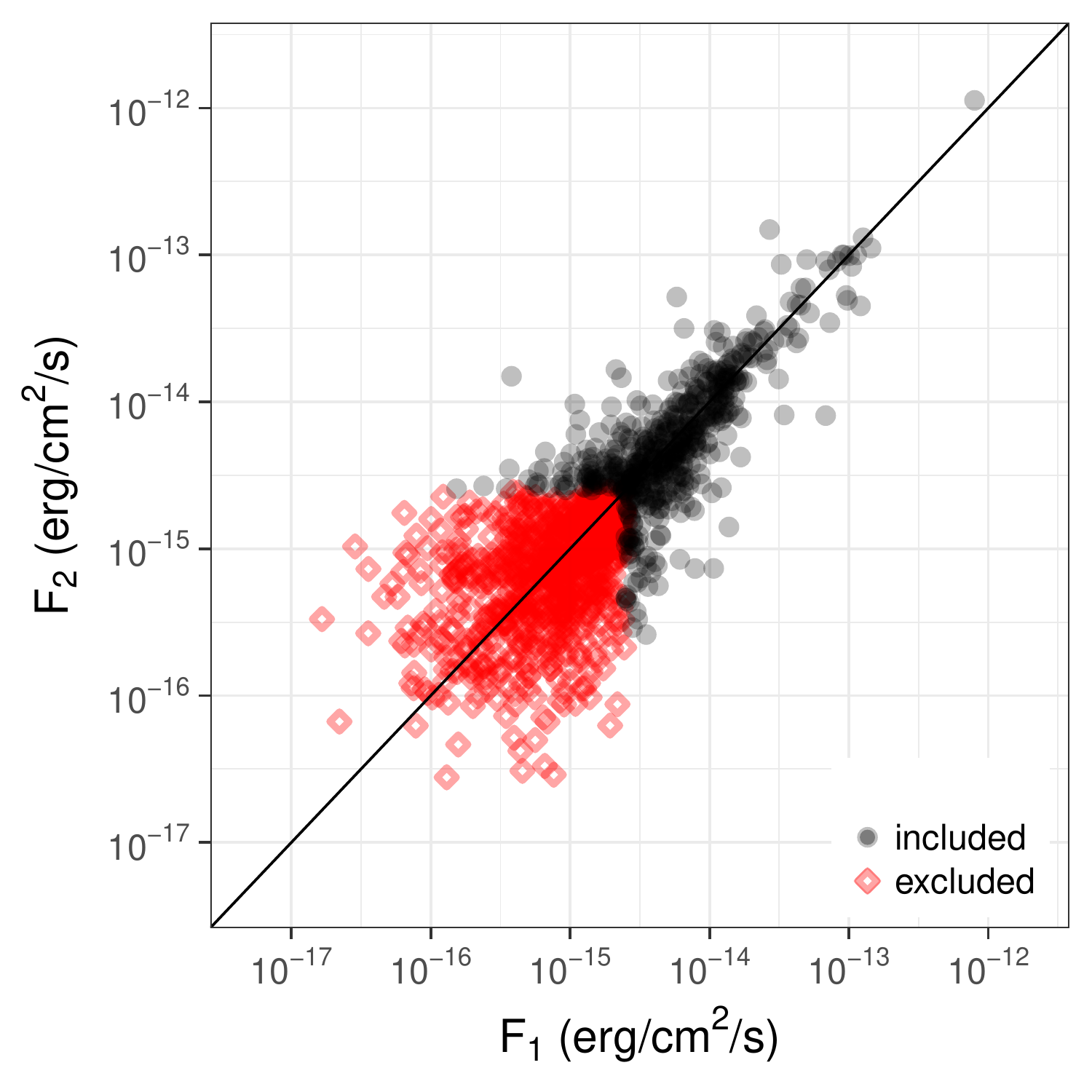}}
\hspace{0.5cm}
\scalebox{0.35}{\includegraphics*{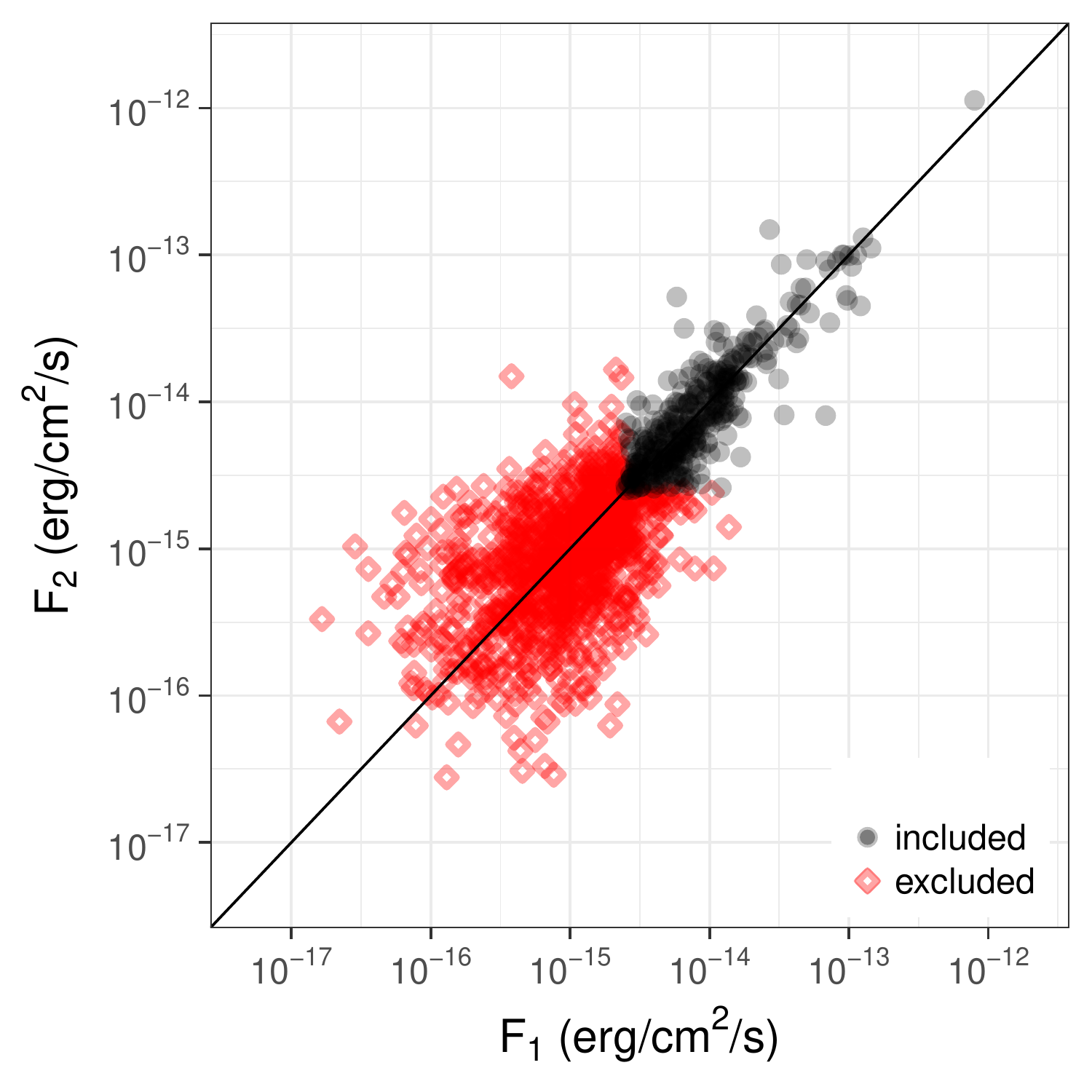}}
\hspace{0.5cm}
\scalebox{0.35}{\includegraphics*{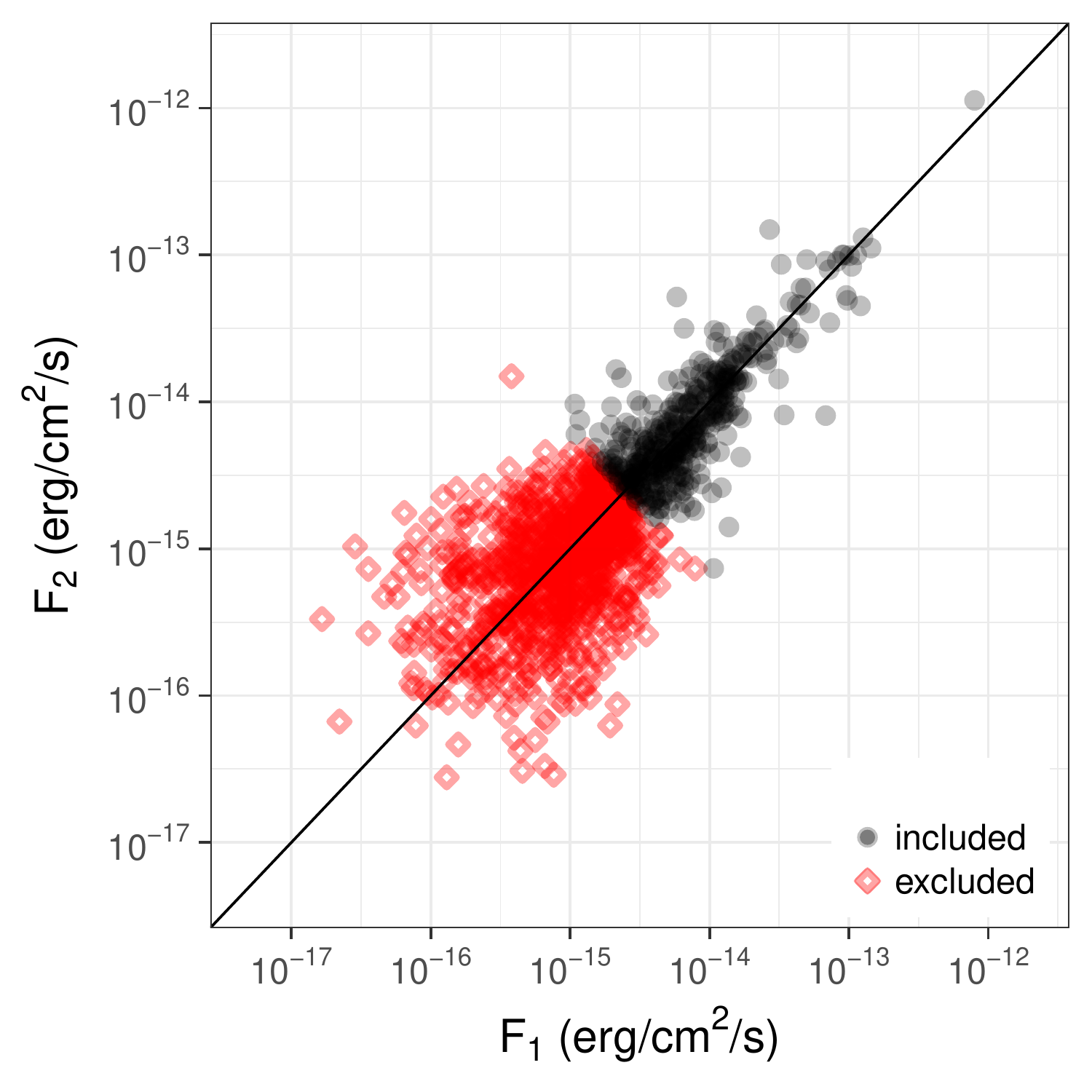}}
\caption[]{\label{fig.selection} Illustration of sample definitions and resulting biases. The points show the measured fluxes in each observation for the full sample, excluding error bars and upper limits for clarity. Filled circles indicate points included by a given sample definition while hollow diamonds indicate excluded points. The left panel illustrates a selection where $(F_1 \text{ OR } F_2) > 2.5 \times 10^{-15}\flux$, which biases the measured $\sigma$ high. The centre panel illustrates a selection where $(F_1 \text{ AND } F_2) > 2.5 \times 10^{-15}\flux$ (biasing the measured $\sigma$ low). The right panel illustrates a selection where $\sqrt{F_1F_2} > 2.5 \times 10^{-15}\flux$, which imposes no bias on $\sigma$.}
\end{center}
\end{figure*}

As an illustration, consider samples defined using a limit in the
observed flux \(F_\text{lim}\). One could define a sample of the AGN for
which (\(F_1 > F_\text{lim}) \text{ OR } (F_2 > F_\text{lim})\). This is
illustrated in the left panel of Fig. \ref{fig.selection} and as is
apparent, this selection results in an overestimate of \(\sigma\) due to
the exclusion of AGN with small flux differences near \(F_\text{lim}\).
An alternative selection of \((F_1 > F_\text{lim}) \text{ AND } (F_2 >
F_\text{lim})\) is illustrated in the central panel of Fig.
\ref{fig.selection}. This is illustrative of a sample where the source
is required to be detected in all observations (for example in a
source catalogue). In this case \(\sigma\) will be \emph{underestimated} due to
the exclusion of AGN with large flux differences close to
\(F_\text{lim}\). In both cases, the slope \(\beta\) of the population
distribution influences the amount of bias on \(\sigma\) since
increasing \(\beta\) increases the density of AGN close to \(F_\text{lim}\),
whence the bias originates.

The selection bias and dependence on \(\beta\) can be avoided entirely by
defining a sample with a flux selection that is orthogonal to the line
of equality of the two fluxes. Since \(\sigma\) is measured in log space,
this selection must also be made in log space, and corresponds to
\(\sqrt{F_1F_2} > F_\text{lim}\) (i.e. the geometric mean of the two
fluxes is greater than \(F_\text{lim}\)). This is illustrated in the
right panel of Fig \ref{fig.selection}, and all of the samples used in
our analysis are defined in this way.

Even using this geometric mean flux limit, one possible source of bias
remains. If \(F_\text{lim}\) were set too low, then the sample would
begin to include false positive sources. Including these sources would
bias \(\sigma\) high, since they would be associated with a background
noise peak in one image and a random background value in the other
image. To minimise this, we set the flux limit to \(F_\text{lim}=2.5
\times 10^{-15}\flux\). This limits the sample to 416 sources of which
408 were detected with a significance \(>3\sigma\) in at least one
observation (the remaining 8 were all detected at \(>2.4\sigma\) in at
least one observation), leading to a highly pure sample.

When computing the geometric mean fluxes for subsample selections, we
used the fluxes measured with \texttt{srcflux}. The mode of the flux
posterior was used for most sources, but in the case of upper limits
(when the mode is zero), the value of the flux upper limit was used
instead. However, no such sources with a flux upper limit exceeded the
flux limits used to define the subsamples for our analysis, so none
were ultimately included in the variability measurements.

\section{Results}
\label{sec:org478de61}
\label{org53f28f3}
We obtained our main results by considering a subsample of AGN for
which the two observations were separated by at least a year, to
reduce sensitivity to short-term variability. This subsample contained
222 AGN, of which 11 were in detected/undetected pairs. We refer to
this as the primary sample.

\begin{figure}
\centering
\includegraphics[angle=0,width=240px]{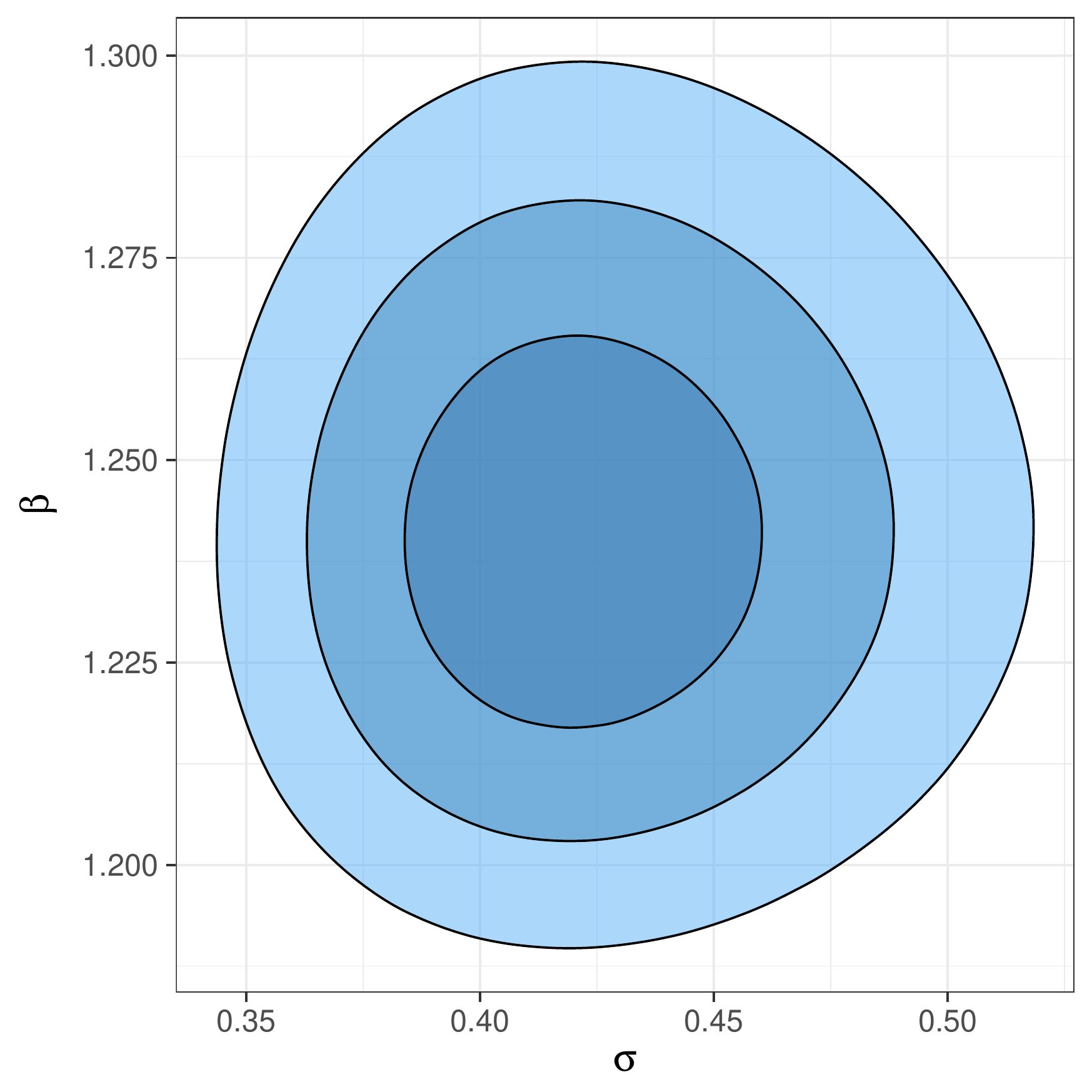}
\caption{\label{fig:orgef70e61}
Constraints on AGN variability \(\sigma\) and \(\lognlogs\) slope \(\beta\) from the primary sample. The shaded regions enclose the \(1\sigma\), \(2\sigma\) and \(3\sigma\) confidence intervals on the parameters.}
\end{figure}

The likelihood in Eq. (\ref{eq.likelihood}) was computed over the
parameter space of \(\sigma\) and \(\beta\) for the primary sample, and
the resulting constraints are shown in Fig. \ref{fig:orgef70e61}.  The best
fitting values were \(\sigma=0.43\pm0.04\) and \(\beta=1.24\pm0.02\).

The constraints on the two parameters are essentially independent with
no degeneracy. For this reason, we fix \(\beta=1.24\) for all subsequent
fits in order to speed up the fitting process. (We will show in
\textsection \ref{orgc1d17df} that this has no impact on the results.) With
\(\beta\) fixed, the constraint on \(\sigma\) for one free parameter was
\(\sigma=0.43\pm0.03\). The results obtained for this and the subsequent
subsamples are summarised in Table \ref{tab:org6b7fee6}.

\begin{table*}
\caption{\label{tab:org6b7fee6}
AGN variability for different subsets of AGN. In all cases, the minimum interval between observations was one year. The first column the allowed range in separation between the sources in the two observations. The second column is the threshold in exposure time that must be exceeded by at least one of the observations. The third column gives the allowed range of the geometric mean flux of the two observations. The fourth column gives the number of AGN in the resulting subsample and the final column gives the constraints on \(\sigma\). For all fits, \(\beta\) was fixed at a value of 1.24.}
\centering
\begin{tabular}{lllrl}
\hline
Separation & Exposure & \(\sqrt{F_1F_2}\) & N & \(\sigma\)\\
(arcsec) & (ks) & \((10^{-15} \flux)\) &  & \\
\hline
\(<1\) & \(>20\) & \(>2.5\) & 222 & \(0.43\pm0.03\)\\
\hline
\(<0.5\) & \(>20\) & \(>2.5\) & 160 & \(0.42\pm0.03\)\\
\(0.5-1\) & \(>20\) & \(>2.5\) & 62 & \(0.45\pm0.05\)\\
\hline
\(<1\) & \(>85\) & \(1.0-2.5\) & 64 & \(0.53\pm0.07\)\\
\hline
\end{tabular}
\end{table*}

In order to test the effect of the matching radius on the measured
variability, we restricted the sample to the 165 AGN in the primary
sample with a matching radius of \(<0.5\arcsec\). For this subset, the
best fitting variability was \(\sigma=0.42\pm0.03\), fully consistent
with the previous measurement. Meanwhile, for the 65 AGN with matching
radii between \(0.5\arcsec\) and \(1\arcsec\), the variability was
\(\sigma=0.45\pm0.05\). We thus conclude that using a \(1\arcsec\)
matching radius has no impact on our results.

Finally, we also investigated if there was evidence for \(\sigma\) being
different for lower flux AGN. To do this we selected AGN for which at
least one of the two observations had an exposure of at least \(85\ks\)
(chosen to give a reasonable number of deep pointings). We then
selected sources with geometric mean fluxes of between \(1 \times
10^{-15}\flux\) and \(2.5 \times 10^{-15}\) to give a low-flux sample of
64 AGN (all but one detected at \(3\sigma\)) that had no overlap with
the primary sample. For this low-flux subsample, the best fitting
variability was \(\sigma=0.53\pm0.07\). This is not a very significant
difference from the value of \(\sigma=0.43\pm0.03\) measured for the
higher flux sources, but could be indicative of a larger variability
at lower fluxes.

\subsection{Variability on different timescales}
\label{sec:org8c73848}
\begin{figure}
\centering
\includegraphics[angle=0,width=240px]{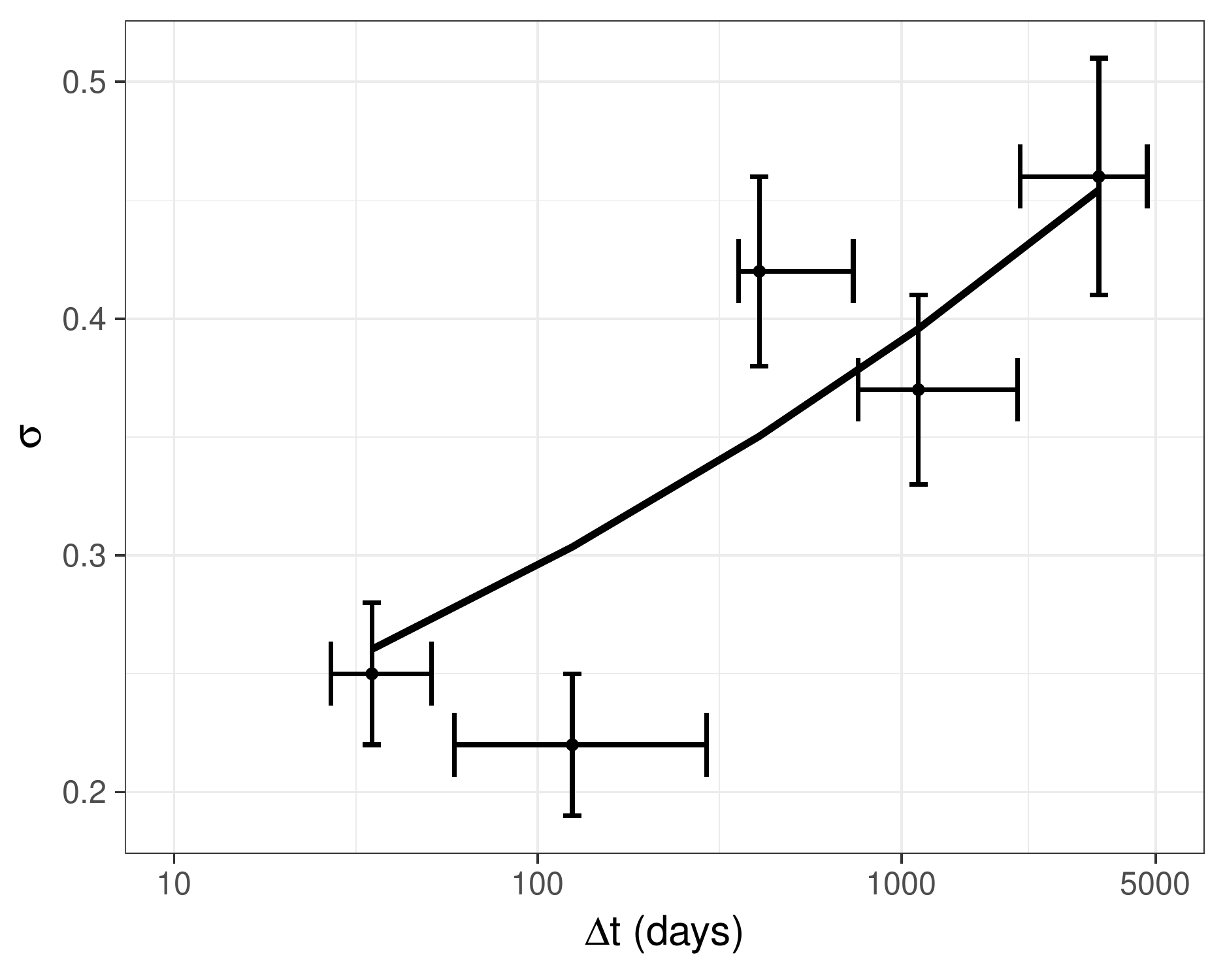}
\caption{\label{fig:org8f56a69}
Constraints on AGN variability \(\sigma\) for subsamples grouped by the interval between observations \(\Delta t\) in the observer's frame. The horizontal bars span the range of \(\Delta t\) in each subsample with the point marking the median \(\Delta t\) for each subsample. The solid line shows the variability as a function of \emph{rest frame} time interval inferred from the structure function measurements of \citet{vag16}.}
\end{figure}

Next, we measured the variability as a function of the (observer's
frame) time difference \(\Delta t\) between observational epochs. For
this we used the same flux selection as the primary sample, but
defined five subsamples of AGN with ranges in \(\Delta t\) chosen to
give approximately equal numbers of AGN in each subsample. The
subsamples were then modelled as before to determine the
characteristic flux variability on those different timescales. The
resulting constraints on \(\sigma\) are shown in Table \ref{tab:org9ab2af8} and
plotted in Fig. \ref{fig:org8f56a69}. Also shown in Fig. \ref{fig:org8f56a69} is the trend of
variability with \emph{rest frame} time interval inferred from the
structure function analysis of a large sample of AGN in \citet{vag16}.
This comparison is discussed in \textsection \ref{org20143b9}.

\begin{table}
\caption{\label{tab:org9ab2af8}
AGN variability for subsets of observations separated by different intervals in the observer's frame. The first column gives the range in intervals between observations and the second gives the median value. The third column gives the number of AGN in the resulting subsample and the final column gives the constraints on \(\sigma\). For all fits, \(\beta\) was fixed at a value of 1.24.}
\centering
\begin{tabular}{lrrl}
\hline
\(\Delta t\) & Median \(\Delta t\) & N & \(\sigma\)\\
(days) & (days) &  & \\
\hline
\(27-51\) & 35 & 95 & \(0.25 \pm  0.03\)\\
\(59-291\) & 124 & 88 & \(0.22 \pm  0.03\)\\
\(357-737\) & 407 & 75 & \(0.42 \pm  0.04\)\\
\(761-2085\) & 1114 & 78 & \(0.37 \pm  0.04\)\\
\(2123-4743\) & 3490 & 80 & \(0.46 \pm  0.05\)\\
\hline
\end{tabular}
\end{table}

\section{Discussion}
\label{sec:org4746ded}
\label{orga29cc76}
\subsection{Verification of methodology}
\label{sec:orgedd0881}
\label{orgc1d17df} Our AGN variability model was tested on simulated data
in order to verify that the variability could be recovered accurately.
A synthetic dataset was generated by sampling a large number of fluxes
from a reference \(\lognlogs\) distribution, with each representing the
mean flux of a different synthetic source. The variability of each
synthetic source was then modelled as a lognormal distribution with
the mean flux for that source and with a constant value of \(\sigma\)
common to all sources. For each synthetic source, a pair of fluxes
were then sampled from its lognormal distribution to represent two
observations of that flux. For each pair of fluxes a random real
source was chosen from the list of all sources detected in our
\emph{Chandra} data (i.e. prior to any filtering), and the pair of
synthetic fluxes were assigned the source and background areas,
exposures, effective areas and background counts of the two observations
of the real source. Finally, these properties were used to compute the
total intensity in the source aperture for each source, and the total and
background counts for each source were then sampled from Poisson
distributions with the appropriate rates. This method produced a large
number of synthetic observations of a realistic population of AGN with
observational characteristics the represent the range of data quality
in the real data.

Following this method, mock samples could be generated with different
characteristics and different selection functions applied to test our
model. In all cases, when the geometric mean flux selection was used,
the input variability was recovered accurately. For example, we
generated a population of AGN following the \(0.5-2\keV\), broken
power-law \(\lognlogs\) distribution of \citet{leh12}, with slopes
\(\beta_1=1.49\) and \(\beta_2=2.48\) at fluxes below and above \(6 \times
10^{-15}\flux\) respectively. The geometric mean flux limit of
\(\sqrt{F_1F_2} > 2.5 \times 10^{-15}\flux\) was then applied to this
synthetic sample and a random subset of 1000 pairs of synthetic
observations was selected. We then modelled this using the methodology
used for our main results, with \(\beta\) fixed at 1.24, and fitting only
for \(\sigma\). For an input variability of \(\sigma=0.5\) our method
recovered \(\sigma=0.50\pm0.02\). This demonstrates that our method is
robust, and due to the geometric mean selection, is insensitive to
the modelling of the \(\lognlogs\) distribution.

In the cases where the OR or AND sample selections discussed above
were applied to the synthetic data, the recovered \(\sigma\) was biased
by \(\approx15\%\) high and low respectively.

\subsection{Evaluation of our model}
\label{sec:org080b594}
\begin{figure}
\centering
\includegraphics[angle=0,width=240px]{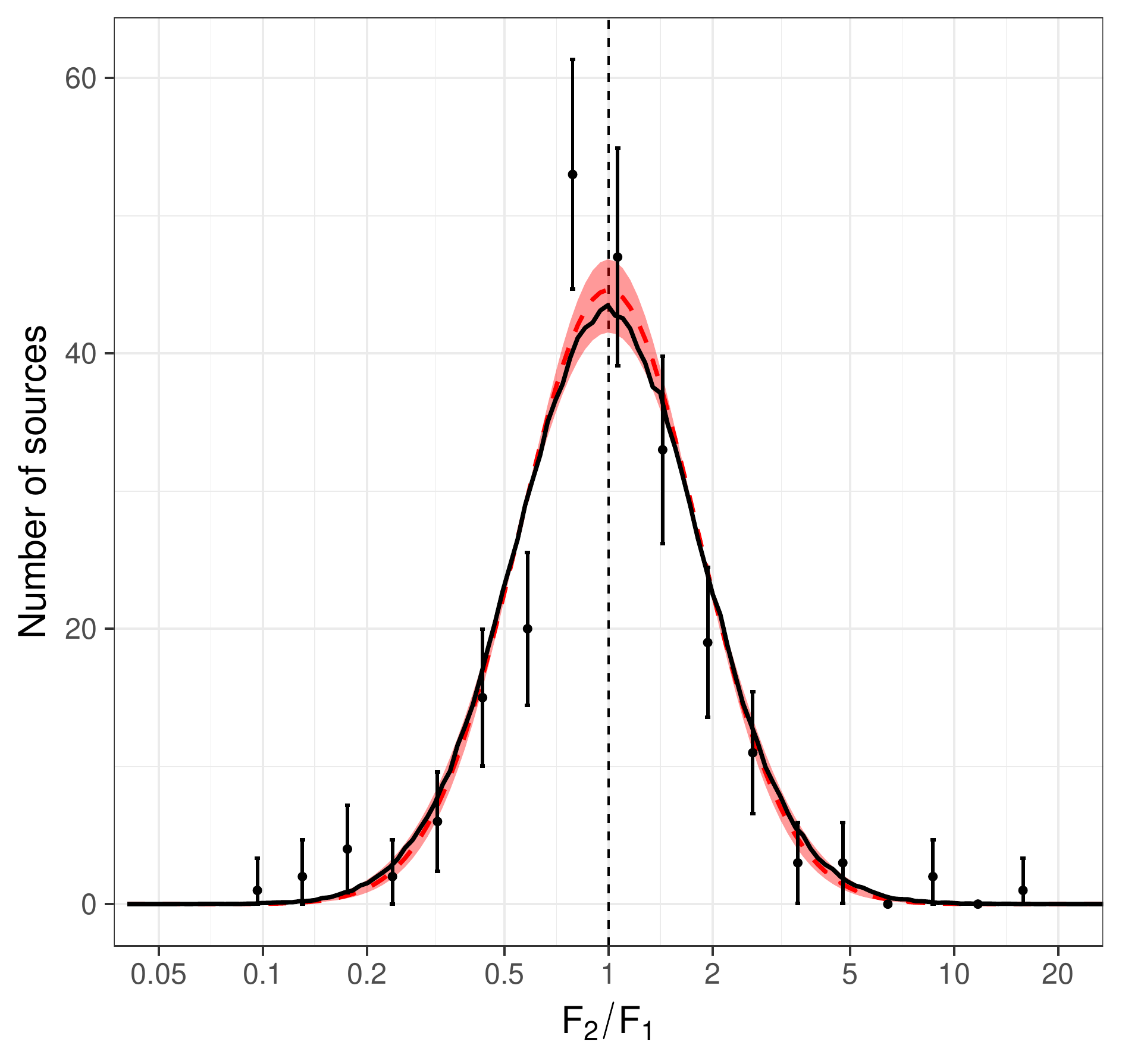}
\caption{\label{fig:orgd2c238e}
The distribution of the ratios of the observed fluxes for the primary sample are compared with the best fitting model. The data points show the number of sources in bins of flux ratio of constant width. The error bars show the Poisson uncertainty on the number of sources in each bin, using the Gaussian approximation of \citet{geh86}. The solid black line shows the distribution of flux ratios derived from a synthetic sample of sources generated from our best fitting model. The dashed red line and shaded \(1\sigma\) error envelope show the lognormal distribution with the best fitting variability of \(\sigma=0.43\pm0.03\) multiplied by a factor of \(\sqrt{2}\). As discussed in the text, this factor accounts for the uncertainty on the mean of a pair of observations. The small difference between the two curves is due to the inclusion of statistical noise and sample selection effects in the curve derived from the synthetic sample.}
\end{figure}

In the previous section we demonstrated that our model can
accurately recover the true variability of realistic synthetic data.
In this section we assess how well our model describes the observed
data. To do this, we computed the ratio of the observed fluxes for
each source in the primary sample (\(F_1/F_2\)). These are plotted in
Fig. \ref{fig:orgd2c238e}, and form the basis of a quantitative comparison
with our model.

According to our model, the two observations of an AGN have fluxes
\(F_1,F_2\) that are drawn from a lognormal distribution with a given
\(\sigma\) (\(\sigma=0.43\) in the case of the primary sample). If the mean,
\(\log(F)\), of the distribution were known, then the ratio of \(F_1/F\)
or \(F_2/F\) would be lognormally distributed with a mean of one and
standard deviation \(\sigma\). However, the ratio of pairs of independent
measurements \(F_2/F_1\) will be lognormally distributed with a mean of
one but with a standard deviation of \(\sqrt{2}\sigma\). As expanded
upon in the next section, the \(\sqrt{2}\) factor arises as \(F_1\) and
\(F_2\) are both independent samples from the flux distribution. This
model is plotted in Fig. \ref{fig:orgd2c238e}, and appears to be a very
good description of the data.

However, with this analytic form we neglect the photon counting noise
on the individual flux measures, which would broaden out the
distribution of flux ratios compared to the lognormal model. To model
this effect we generated \(10^6\) synthetic pairs of observations of AGN
following the method of \textsection \ref{orgc1d17df}, but using the
best fitting model to our primary subsample (i.e. \(\sigma=0.43\), and a
single power-law \(\lognlogs\) with \(\beta=1.24\)). The geometric mean flux
limit of \(\sqrt{F_1F_2} > 2.5 \times 10^{-15}\flux\) was then applied
to this synthetic data to match the primary sample selection, and the
ratios of \(F_1/F_2\) were computed for each AGN.

The distribution of flux ratios predicted by the best fitting model is
plotted in Fig. \ref{fig:orgd2c238e}. As expected, this is slightly
broadened compared with the intrinsic lognormal distribution. Visually
there is a good agreement with the observed flux ratios in this binned
representation. The unbinned distribution of observed flux ratios was
compared with that of the synthetic sample from the best fitting model
using a two-sided Kolmogorov-Smirnov test. This gave a p value of
0.74, meaning that the data cannot rule out the null hypothesis that
both the observed and synthetic samples are drawn from the same parent
distribution. We thus conclude that our model is a good description of
the data.

\subsection{Comparison with other work}
\label{sec:org6392466}
\label{org20143b9}
Our analysis is quite distinct from previous work on the long term
variability of AGN, in that it is not focussed at understanding the
intrinsic properties of the AGN, but instead on equipping the observer
with the knowledge needed to mitigate their impact on other targets of
interest. As such we work exclusively in the observer's frame rather
than the rest frame of the source, and do not separate out AGN into
samples by redshift or luminosity.

Our analysis method has some advantages over other work. Firstly, we
define a sample selection that avoids the biases introduced when
applying an independent selection threshold to each observation.
Secondly, we use the Poisson likelihood of the observed counts, which
removes the approximation of Gaussian statistics and naturally includes
non-detections and upper limits.

In the literature, two main approaches have been used to measure the
characteristic variability of samples of AGN: the normalised excess
variance (NXS) and the structure function (SF). Both quantities are
defined precisely in \citet{vag16} but they can be described as
follows. The NXS is the variance of the flux distribution of a
particular source once measurement errors have been subtracted,
normalised to the mean flux of that source. The SF is the root mean
square of the difference in log flux between pairs of observations,
where the average is taken over sources with approximately the same
time interval between observations, and the average statistical noise
is subtracted.

\citet{alm00} used a variation on the NXS method to measure the
variability in 86 quasars on timescales of up to 14 days. Their method
is not directly comparable as they utilised 16-26 flux measurements
per source, spread over the 14 day period, but the average variability
of \(\approx20\%\) that they found is similar to the value we find for
the shortest time intervals we sampled.

\citet{mat07} used the same approach as \citet{alm00} to measure
variability in AGN observed in 16 observations of the Lockman Hole
field with \(\XMM\) spread over 2 years. The average fractional
variability for that sample was \(0.22\pm0.01\), which is smaller than
the values of \(\approx0.4\) that we find on timescales longer than
about a year. However, once again a direct comparison is difficult as
the \citet{mat07} measurement comes from multiple fluxes spread over 2
years while ours comes from flux pairs separated by a given interval.
The mean interval between observations for the \citet{mat07} value was
about 40 days, so a better comparison may be with the \(\approx25\%\)
variability we find on that timescale.

\citet{vag16} investigated the variability in a sample of 2700 AGN
observed multiple times with \(\XMM\), and with known redshifts using
both the NXS and SF methods. Their SF measurements are quite
comparable with our measurement of characteristic variability as they
are determined from pairs of fluxes separated by \(\Delta t\), and we
thus convert their SF to a fractional variability using their
equation 6.

The resulting trend in fractional variability with observation
interval for the whole \citet{vag16} sample is shown in Fig. \ref{fig:org8f56a69},
and the agreement with our measurements is very good. This is despite
some significant differences between these works. In particular, the sample
definition of \citet{vag16} requires the sources to be detected in all
observations, which should lead to an underestimate of the true
variability, while our analysis uses the observation interval in the
observer's frame. The latter effect would blur out pairs of
observations with the same rest frame interval into a range of
observer's frame intervals, flattening the slope of any trend between
\(\sigma\) and \(\Delta t\). The good agreement between the two sets of
results suggests that neither of these effects are large.

Overall, we conclude that while previous measurements using the NXS
method are not directly comparable, the agreement with our results
seems reasonable. The closest comparison is with the SF measurements
of \citet{vag16}, where the agreement is very good.

\subsection{Forecasting fluxes}
\label{sec:orgc212fa7}
We are now in a position to answer the question posed in this paper.
Given a measurement of the X-ray flux \(F_1\) of an AGN at one epoch,
and with no additional knowledge of its properties, what is our best
estimate of the flux \(F_2\) at a second epoch some years earlier or
later? Under these circumstances, neglecting Eddington bias (see
below), the flux is equally likely to be higher or lower in the second
observation, so our best estimate must be \(F_2=F_1\), but the
uncertainty on this depends on the variability of the source. (Of
course the uncertainty also depends on the statistical noise on each
observation, but we disregard that here to focus on the irreducible
uncertainty due to variability.)

Consider an AGN that has a lognormal variability of its flux about a
mean flux \(\log(F)\), with standard deviation \(\sigma\). In this case, given
\(N\) observations of the source with fluxes \(F_i\), we estimate
\(\log(F)\) as the mean of the \(\log(F_i)\) with a precision of
\(\sigma/\sqrt{N}\). In other words, the probability density for \(F\) is
\begin{align}
P(F|F_i,\sigma) = \dlnorm(\left<\log(F_i)\right>,\sigma/\sqrt{N}).
\end{align}

Now, to predict the flux \(F'\) of this source at the epoch of another
observation, we have to marginalise out the unknown mean \(F\) for which
we know the posterior from the previous \(N\) measurements of the flux.
\begin{align}\label{eq.pfprime}
P(F'|F_i,\sigma) = \int P(F'|F,\sigma)  P(F|F_i,\sigma) \, \diff F
\end{align}
where both of the probabilities on the right hand side are lognormal.
This results in a posterior for \(F'\) that is also lognormal, with
standard deviation
\begin{align}\label{eq.sigmas}
\sigma' & =\sqrt{\sigma^2 + \sigma^2/N} = \sigma\sqrt{1+\frac{1}{N}}
\end{align}

Thus, for a source with one previous flux measurement \(F_1\) we predict
the flux at a second epoch to be \(F_2=F_1\) with an uncertainty on
\(\log(F_2)\) of \(\sigma'=\sqrt{2}\sigma\), as above. For example, based
on the \(\sigma=0.43\) we found for the primary sample, for observations
separated by about \(1-10\) years the uncertainty on a flux prediction
based on a single previous measurement is \(0.43\times\sqrt{2}\) or
approximately \(60\%\).

In the limit of a large number of previous flux measurements, the
uncertainty on the log of the flux at a new epoch tends to \(\sigma\).
In principle this sets the average population variability, \(\sigma\) as
the limiting precision of any flux forecast. However, with many flux
measurements of the same source, the variability of that source could
be constrained, resulting in a more accurate prediction of its flux at
another epoch (with a precision limited by the variability of that
particular source). As illustrated in \ref{fig:org8f56a69}, further improvements
can be gained by scheduling observations with the shortest possible
intervals between them to reduce the overall variability.

Two further effects should also be considered when making flux
predictions. Firstly, if an AGN is discovered in an observation
at some epoch, there is an Eddington bias effect present. Given the
greater number of AGN at low fluxes (as described by the \(\lognlogs\)
distribution), a newly discovered AGN is more likely to be a low flux
AGN in a high flux state than vice-versa. We would thus expect the AGN
to be more likely to have a lower flux at another epoch, regressing to
the mean. This can be modelled by including the population
distribution \(P(F)\) in equation \ref{eq.pfprime}:
\begin{align}
P(F'|F_i,\sigma) = \int P(F'|F,\sigma)  P(F|F_i,\sigma) P(F) dF
\end{align}
where \(P(F)\) could be e.g. a single or double power law. Secondly, in
order to forecast the \emph{observed} value of \(F'\), it is further
necessary to model the statistical noise on the observation, using
knowledge of the appropriate instrumental and observational
characteristics.

\begin{figure}
\centering
\includegraphics[angle=0,width=240px]{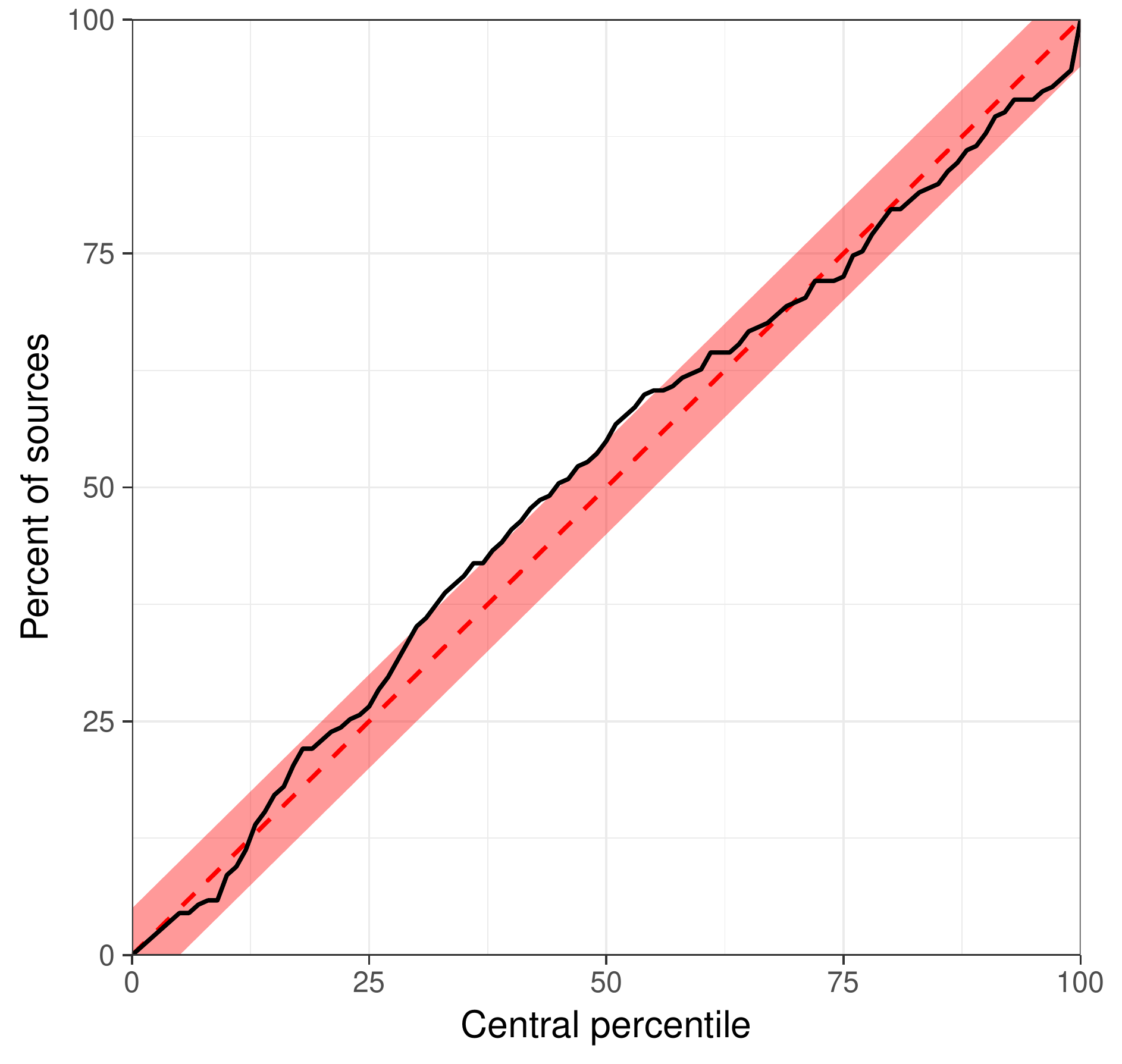}
\caption{\label{fig:orgcca9224}
Percentage of sources in the primary sample whose predicted flux at epoch 2 fell within a given central percentile of a lognormal distribution with \(\sigma=0.43\sqrt{2}\) centred on the flux at epoch 1. The solid line shows the data, while the dashed line shows the expected one-to-one correspondence if the model perfectly predicted the data. The shaded region encloses plus or minus five percentage points around the model prediction.}
\end{figure}

We investigated how well our model predicted \(F_2\) given \(F_1\) for our
primary sample. Neglecting the effects of statistical noise and
Eddington bias, our best estimate of the flux at the second epoch is
\(F_2=F_1\pm\sqrt{2}\sigma\), where the uncertainty is a lognormal
distribution. We then calculated the percentage of sources for which
\(F_2\) fell within a given central percentile of this lognormal
distribution of \(F_1\). For example, if the model gives good
predictions then we expect that about \(68\%\) of the time, the
predicted \(F_2\) values will be within the central 68th percentile of a
lognormal distribution centred on \(F_1\) (i.e. within the \(1\sigma\)
error on the forecast). The resulting forecasts are shown in Fig.
\ref{fig:orgcca9224}, which shows that the precision on the flux forecast is
good to within 5 percentage points for all confidence intervals.

Our formalism and results are applicable for the scenario in which the
redshift of the AGN is unknown. If the redshift and hence luminosity
is known, then the accuracy with which the flux at a second epoch can
be predicted is significantly improved. This is because the
variability in X-ray luminosity of AGN is a function of luminosity,
such that lower-luminosity AGN show larger fractional variability
\citep[e.g][]{pap93a,vag16}. For example, \citet{vag16} find the
fractional variability for observations separated by \(\Delta t = 1000\)
days (rest frame) to range from \(\sigma\approx0.55\) for their least
luminous AGNs (\(10^{43} \erg\ps - 10^{43.5}\erg\ps\) in the
\(0.5-4.5\keV\) band) down to \(\sigma\approx0.25\) for their most
luminous AGNs (\(10^{45} \erg\ps - 10^{45.5}\erg\ps\) in the
\(0.5-4.5\keV\) band). This range in luminosity variability is naturally
included in the average flux variability that we measure. However, if
the redshift and hence luminosity of an AGN were known, then the
results of \citet{vag16}, or similar studies, could be used to
estimate a more accurate variability for that particular AGN.

\subsection{Applications}
\label{sec:org60309c8}
This measurement of the characteristic X-ray variability of AGN has
important consequences for scenarios when the flux of an AGN at one
epoch must be inferred from its flux at a second. Given the uniquely
high angular resolution of \emph{Chandra} compared to other X-ray
observatories, the most common such scenario will continue to be the
use of \emph{Chandra} to constrain point source contributions to
observations made with other observatories.

A key example is the use of \emph{Chandra} to support deep \(\XMM\)
observations of distant galaxy clusters, where \(\XMM\) cannot resolve
out the emission from projected AGN, biasing measurements of the
temperature and luminosity of the cluster \citep[e.g.][]{hil10}. As
demonstrated by \citet{hil10}, the effect of such contamination can be
mitigated by jointly modelling \emph{Chandra} data to constrain the AGN
flux. In doing this, the uncertainty on the flux of the source due to
its variability between epochs should be included. For example, if the
interval between observations were a year or more, then the
uncertainty on the flux prediction based on a single measurement is
\(\approx 60\%\). This could be modelled with a lognormal prior with
\(\sigma=0.6\) on the flux (or normalisation of the relevant model
component) of the AGN at the epoch of the \(\XMM\) observation, centred
on the flux measured with \emph{Chandra}. This approach will also be
relevant for observations of distant clusters with ATHENA, whose
angular resolution will be poorer than that of \emph{Chandra}. Where
possible, the interval between observations should be minimised to
reduce the average variability. The same approach can be used in
studies of cluster outskirts, which may optimally combine \emph{Chandra}
measurements of the point source population with \emph{Suzaku}'s detection
of the ICM \citep[e.g.][]{tho16}.

The presence of AGN can also bias the detection of galaxy clusters in
X-ray surveys such as XXL \citep{pie16}, XCS \citep{meh12} or the
upcoming eROSITA survey \citep{pil12,mer12,pil18,cle18}. In all cases
the relatively poor angular resolution of the survey data may lead to
AGN being misclassified as clusters, or boost the detection
probability of clusters by enhancing their surface brightness
(clusters with projected AGN may also be misclassified as pure AGN and
missed by these cluster surveys). The purity of such surveys can be
estimated by short \emph{Chandra} observations of a subset of clusters to
detect the presence of contaminating AGN (Logan et. al., 2018,
submitted). The variability between epochs should then be included
when considering the contaminating flux (or upper limit thereon)
determined from the \emph{Chandra} data.

A further application of our constraints on X-ray variability is to
inform the modelling of stray light in X-ray observations due to
bright point sources outside the field of view. Stray light from such
sources can produce faint, inhomogeneous structure in X-ray images
that could bias studies of low surface brightness emission. A model of
the stray light component could be produced to mitigate this, given an
accurate model of the X-ray optics and knowledge of the fluxes and
positions of sources outside the field of view. The latter could come
from all-sky X-ray survey data. However, in the case that the stray
light signal is dominated by a few bright AGN, their variability
between the epoch(s) of the survey data providing their fluxes and the
observation for which the stray light must be modelled will provide an
irreducible limit on the precision of the stray light modelling.

\section{Summary and conclusions}
\label{sec:org5c4e68d}
\label{orga22dc7d}
We have used pairs of Chandra observations of a large number of AGN to
infer the characteristic X-ray variability of the population on
different timescales. We developed a sample selection method that is
insensitive to biases, and a likelihood model that was able to
precisely recover the variability in realistic simulated data.

For our primary sample of sources observed between around 1 and 15
years apart, the variability of the population is well described by a
lognormal distribution with standard deviation of
\(\sigma=0.43\pm0.03\). We find evidence, in common with other work,
that the variability is smaller on shorter timescales, with
\(\sigma\approx0.25\) for separations of about one month to one year
between observations.

Given a single flux measurement, the best estimate of the flux at a
second epoch is the same as that at the first epoch (neglecting
Eddington bias) with a (lognormal) \(68\%\) confidence interval of
\(\sqrt{2}\sigma\). The factor \(\sqrt{2}\) arises due to the uncertainty
on the mean flux for the source given just one previous measurement.

As a rule of thumb, given the flux of an AGN at one epoch, one can
estimate its flux at a second epoch (more than a year or so earlier or
later) to a precision of about \(60\%\).

This result has applications in a wide range of scenarios where X-ray
fluxes of AGN must be inferred from their values at a different epoch,
and presents a significant source of irreducible uncertainty that
should be taken into account. A useful next step would be to better
constrain the dependence of the variability on interval between
measurements, including longer time intervals then those probed here
as data become available.

\section*{Acknowledgements}
BJM acknowledges support from STFC grant ST/R000700/1. THR acknowledges support by the German Aerospace Agency (DLR) with funds from the Ministry of Economy and Technology (BMWi) through grant 50 OR 1514. This project made extensive use of TOPCAT \citep[Tool for OPerations on Catalogues And Tables][]{tay05}. Preliminary work was done by University of Bristol undergraduate students James Battye, Jane Hesling and Nicholas Henden.

\bibliographystyle{mnras}
\bibliography{clusters}
\end{document}